\documentclass[pdflatex,sn-standardnature,referee]{sn-jnl_arXiv_version}

\usepackage[utf8]{inputenc}
\usepackage{anyfontsize}
\usepackage{pdfpages}

\unnumbered

\usepackage{hyperref}
\usepackage[all]{hypcap} 

\begin{document}

\title[Local Gate Control of Mott MIT in a 2D MOF]{Local gate control of Mott metal-insulator transition in a 2D metal-organic framework}

\author[1,2]{Benjamin Lowe}
\equalcont{These authors contributed equally to this work.}
\author[1,2]{Bernard Field}
\equalcont{These authors contributed equally to this work.}
\author[1,2]{Jack Hellerstedt}
\author[1,2]{Julian Ceddia}
\author[3]{Henry L. Nourse}
\author*[4]{Ben J. Powell}
\author*[2,5]{Nikhil V. Medhekar}
\author*[1,2]{Agustin Schiffrin}
\affil[1]{School of Physics and Astronomy, Monash University, Clayton, Victoria 3800, Australia}
\affil[2]{ARC Centre of Excellence in Future Low-Energy Electronics Technologies, Monash University, Clayton, Victoria 3800, Australia}
\affil[3]{Quantum Information Science and Technology Unit,
Okinawa Institute of Science and Technology Graduate University, Onna-son, Okinawa 904-0495, Japan}
\affil[4]{School of Mathematics and Physics, The University of Queensland, Brisbane, Queensland 4072, Australia}
\affil[5]{Department of Materials Science and Engineering, Monash University, Clayton, Victoria 3800, Australia}
\email{powell@physics.uq.edu.au}
\email{nikhil.medhekar@monash.edu}
\email{agustin.schiffrin@monash.edu}

\abstract{Electron-electron interactions in materials lead to exotic many-body quantum phenomena including Mott metal-insulator transitions (MITs), magnetism, quantum spin liquids, and superconductivity. These phases depend on electronic band occupation and can be controlled via the chemical potential. Flat bands in two-dimensional (2D) and layered materials with a kagome lattice enhance electronic correlations. Although theoretically predicted, correlated-electron Mott insulating phases in monolayer 2D metal-organic frameworks (MOFs) with a kagome structure have not yet been realised experimentally. Here, we synthesise a 2D kagome MOF on a 2D insulator. Scanning tunnelling microscopy (STM) and spectroscopy reveal a MOF electronic energy gap of \ensuremath{\mathord{\sim}}200 meV, consistent with dynamical mean field theory predictions of a Mott insulator. Combining template-induced (via work function variations of the substrate) and STM probe-induced gating, we locally tune the electron population of the MOF kagome bands and induce Mott MITs. These findings enable technologies based on electrostatic control of many-body quantum phases in 2D MOFs.}


\maketitle

Strong electronic correlations arise in a material at specific electron fillings of its bands, provided that the on-site Coulomb repulsion (characterised by the Hubbard energy, $U$) is of the order of, or larger than, the bandwidth, $W$. These electronic correlations can result in a wide range of exotic many-body quantum phases. Examples include correlated insulating phases, quantum spin liquids, correlated magnetism, and superconductivity -- phenomena which have been realised in monolayer transition metal-dichalcogenides \cite{nakata_monolayer_2016,liu_monolayer_2021, lin_scanning_2020,chen_strong_2020,ruan_evidence_2021,chen_evidence_2022}, twisted few-layer graphene \cite{cao_unconventional_2018,park_tunable_2021}, inorganic kagome crystals \cite{zheng_emergent_2022,ye_massive_2018,yin_quantum-limit_2020}, and organic charge transfer salts \cite{powell_quantum_2011,kanoda_mott_2011,miksch_gapped_2021}. 

Tuning of the chemical potential via electrostatic gating can allow for control over such band electron filling, enabling reversible switching between correlated phases \cite{cao_unconventional_2018}. This makes these systems amenable to integration as active materials in voltage-controlled devices, offering enticing prospects for applications in electronics, spintronics, and information processing and storage \cite{powell_quantum_2011,shao_recent_2018}.

 Two-dimensional (2D) materials have emerged as particularly promising candidates for realising strongly correlated phenomena as the absence of interlayer hopping and screening can contribute to decreasing $W$ and increasing $U$ \cite{chen_strong_2020}. Additionally, some 2D crystal geometries -- such as the kagome structure -- give rise to intrinsic flat electronic bands \cite{balents_superconductivity_2020,neupert_charge_2022}. When these extremely narrow bands are half-filled, even weak Coulomb repulsion can open an energy gap and give rise to a Mott insulating phase \cite{powell_quantum_2011}. Away from half-filling, the gap closes and the system becomes metallic.

Metal-organic frameworks (MOFs) are a broad class of materials whose properties are highly tunable through careful selection of constituent organic molecules and metal atoms \cite{cui_metalorganic_2016}. There has been growing interest in 2D MOFs for their electronic properties \cite{zhang_intrinsic_2016, jin_large-gap_2018, jiang_exotic_2021}. In particular, layered 2D MOF structures have been recently shown to host strongly correlated superconductivity \cite{takenaka_strongly_2021}. Monolayer 2D MOFs have attracted attention for their magnetism \cite{yamada_designing_2017,zhang_two-dimensional_2019,kumar_manifestation_2021}, including ferromagnetism resulting from exchange interaction between unpaired metal centre electrons \cite{lobo-checa_ferromagnetism_2024}. Despite theoretical predictions \cite{fuchs_kagome_2020, nourse_multiple_2021}, correlated Mott phases have not yet been realised experimentally in monolayer 2D MOFs, however. 

Here, we demonstrate local electrostatic control over a Mott metal-insulator-transition (MIT) in a single-layer 2D kagome MOF, in excellent agreement with theoretical predictions.

\section{Results}
\subsection{A monolayer 2D metal-organic framework on an atomically thin insulator}

We synthesised the monolayer MOF -- consisting of 9,10-dicyanoanthracene (DCA) molecules coordinated to copper (Cu) atoms -- on monolayer hexagonal boron nitride (hBN) on Cu(111) (see Methods for sample preparation). A scanning tunnelling microscopy (STM) image of a crystalline single-layer MOF domain grown seamlessly across the hBN/Cu(111) substrate is shown in Fig. \ref{1}a. We observe some defects within the MOF domain, as well as some DCA-only regions (discussed in Supplementary Note 11). The long-range modulation of the MOF STM apparent height follows the hBN/Cu(111) moir\'{e} pattern, which arises due to mismatch between the hBN and Cu(111) lattices (giving rise to pore, P, and wire, W, regions -- see upper inset) \cite{joshi_boron_2012,zhang_tuning_2018,auwarter_hexagonal_2019}. This moir\'e pattern has been shown to affect the electronic properties of adsorbates \cite{auwarter_hexagonal_2019}, including one previous example of a MOF \cite{urgel_controlling_2015}.
 
 The MOF is characterised by a hexagonal lattice (lattice constant: 2.01 $\pm$ 0.06 nm), with a unit cell including two Cu atoms (honeycomb arrangement; bright protrusions in Fig. \ref{1}b) and three DCA molecules (kagome arrangement, with protrusions at both ends of anthracene backbone in STM image in Fig. \ref{1}b), similar to previous reports \cite{yan_synthesis_2021,yan_two-dimensional_2021,kumar_manifestation_2021,hernandez-lopez_searching_2021}.

We calculated the band structure of this monolayer DCA\textsubscript{3}Cu\textsubscript{2} MOF on hBN/Cu(111) by density functional theory (DFT; with $U = 0$); Fig. \ref{1}d. Projection of the Kohn-Sham wavefunctions onto MOF states shows the prototypical kagome energy dispersion with two Dirac bands and a flat band, consistent with prior theoretical calculations for the freestanding MOF \cite{zhang_intrinsic_2016,fuchs_kagome_2020, field_correlation-induced_2022}. This near-Fermi band structure has predominantly molecular DCA character, and is well described by a nearest-neighbour tight-binding (TB) model (see corresponding density of electronic states, DOS, as a function of energy $E$ in Fig. \ref{1}e \cite{field_correlation-induced_2022}). The hBN monolayer, a 2D insulator with a bandgap $>5$ eV \cite{auwarter_hexagonal_2019}, prevents electronic hybridization between the underlying Cu(111) surface and the 2D MOF \cite{auwarter_hexagonal_2019}. This allows the MOF to preserve its 
 intrinsic electronic properties, in contrast to previous findings on metal surfaces \cite{pawin_surface_2008,hernandez-lopez_searching_2021,kumar_manifestation_2021,field_correlation-induced_2022}. These $U = 0$ calculations predict that the MOF on hBN/Cu(111) is metallic, with some electron transfer from substrate to MOF leading to the chemical potential lying above the Dirac point, close to half-filling of the three kagome bands (Fig. \ref{1}d). 

Strong electronic interactions have been theoretically predicted in DCA\textsubscript{3}Cu\textsubscript{2} \cite{fuchs_kagome_2020}, and a signature was recently detected experimentally \cite{kumar_manifestation_2021}. We therefore calculated the many-body spectral function $A(E)$ -- analogous to the DOS in the non-interacting regime -- of the free-standing MOF via dynamical mean-field theory (DMFT). In contrast to the TB model or DFT, DMFT explicitly captures local electronic correlations caused by the Hubbard energy $U$ (see Methods) \cite{georges_dynamical_1996,kotliar_strongly_2004,vollhardt_dynamical_2012,adler_correlated_2018}. In Fig \ref{1}e, 
$A(E)$ is shown for $U$ = 0.65 eV (consistent with previous experimental estimates \cite{yan_synthesis_2021}) and for a chemical potential which matches the DFT-predicted occupation of the kagome bands for the MOF on hBN/Cu(111) (see Supplementary Note 2 and Supplementary Note 23 for further DMFT calculations, including temperature dependence).
We observe two broad peaks (lower and upper Hubbard bands) separated by an energy gap of \ensuremath{\mathord{\sim}}200 meV, dramatically different from the non-interacting kagome DOS, and indicative of a Mott insulating phase \cite{ohashi_mott_2006, fuchs_kagome_2020}. 

\subsection{Observation of $\sim$200 meV Mott energy gap}

To experimentally probe the electronic properties of DCA\textsubscript{3}Cu\textsubscript{2}/hBN/Cu(111), we conducted differential conductance ($\textrm{d}I/\textrm{d}V$) scanning tunnelling spectroscopy (STS); $\textrm{d}I/\textrm{d}V$ is an approximation of the local DOS [$A(E)$] in the non-interacting (interacting, respectively) picture.
We performed STS at the ends of the DCA anthracene moiety and at the Cu sites of the MOF -- locations where we expect the strongest signature of the kagome bands based on the spatial distribution of the orbitals that give rise to these bands \cite{yan_synthesis_2021,yan_two-dimensional_2021,kumar_manifestation_2021}. These spectra (Fig. \ref{2}a), taken at a pore region of the hBN/Cu(111) moir\'{e} pattern, both show broad peaks at bias voltages $V_{\textrm{b}}\approx-0.2$ and 0.2 V. In a bias voltage window of $\sim$0.2 V around the chemical potential $E_{\textrm{F}}$ ($V_{\textrm{b}}=0$), the $\textrm{d}I/\textrm{d}V$ signal is low, significantly smaller than that for bare hBN/Cu(111).

STM images acquired within this low-$\textrm{d}I/\textrm{d}V$ bias voltage window (Fig. \ref{2}b, c) show mainly the topography of the MOF, with the molecules appearing as ellipses of uniform intensity and the Cu atoms as weak protrusions. Outside the low-$\textrm{d}I/\textrm{d}V$ bias voltage window ($\vert V_{\textrm{b}} \vert > 200$ mV), Cu sites and the ends of the DCA anthracene moieties appear as bright protrusions (Fig. \ref{2}d, e), similar to the spatial distribution of the electronic orbitals of the DCA\textsubscript{3}Cu\textsubscript{2} MOF associated with the near-Fermi kagome bands (right inset of Fig. \ref{2}d; see Supplementary Figs. 11, 12 for more $V_{\textrm{b}}$-dependent STM images and $\mathrm{d}I/\mathrm{d}V$ maps) \cite{yan_synthesis_2021,yan_two-dimensional_2021,kumar_manifestation_2021}.

This suggests that the $\textrm{d}I/\textrm{d}V$ peaks at $\vert V_{\textrm{b}} \vert \approx0.2$ V in Fig. \ref{2}a are related to intrinsic MOF electronic states near $E_{\textrm{F}}$, with the low-$\textrm{d}I/\textrm{d}V$ bias voltage window of $\sim$0.2 V around $E_{\textrm{F}}$ representing an energy gap, $E_\mathrm{g}$, between these states. This is consistent with in-gap topographic STM imaging in Fig. \ref{2}b, c \cite{repp_molecules_2005}. These $\textrm{d}I/\textrm{d}V$ peaks cannot be attributed to inelastic tunnelling (e.g., MOF vibrational modes) as they are not always symmetric about $E_{\textrm{F}}$ (see Fig. \ref{3}). The gap is much larger than that predicted from spin-orbit coupling in such a system \cite{zhang_intrinsic_2016}, and the monocrystalline growth of the MOF domains (Fig. \ref{1}a) makes a large disorder-related gap unlikely. Furthermore, the gap is inconsistent with DFT and TB calculations (Fig. \ref{1}d, e).

The MOF spectra in Fig. \ref{2}a strongly resemble the DMFT-calculated spectral function $A(E)$ in Fig. \ref{1}e, including an energy gap of the same magnitude between two similar peaks (Supplementary Fig. 9). This suggests that these $\mathrm{d}I/\mathrm{d}V$ spectra are hallmarks of a Mott insulator.  

\subsection{Template-assisted Mott metal-insulator transition}\label{tem_ind_gat}

We further measured $\mathrm{d}I/\mathrm{d}V$ spectra at Cu sites of the DCA\textsubscript{3}Cu\textsubscript{2} MOF across the hBN/Cu(111) moir\'{e} pattern (Fig. \ref{3}a, b).
In Fig. \ref{3}b, $E_\mathrm{g}$ is centred symmetrically about $E_\mathrm{F}$ for spectra taken in the middle of a pore region, while those taken closer to the wire region show the Hubbard bands shifting upwards in energy (lowering the barrier to creation of a hole). 
At the centre of the wire region, the gap at the Fermi level vanishes with a clear increase in Fermi-level $\mathrm{d}I/\mathrm{d}V$ signal (Fig. \ref{3}e). The same behaviour was observed for DCA lobe sites of the MOF (see Supplementary Note 14).

The hBN/Cu(111) moir\'{e} pattern consists of a modulation of the local work function $\Phi$ (with little structural corrugation), where the quantity $\Delta \Phi = \Phi_{\textrm{wire}} - \Phi_{\textrm{pore}}$ depends on the period of the moir\'{e} superstructure, $\lambda$ \cite{joshi_boron_2012,zhang_tuning_2018,auwarter_hexagonal_2019}. For the hBN/Cu(111) domain in Fig. \ref{3}a with $\lambda \approx 12.5$ nm, $\Delta \Phi \approx0.2$ eV (Fig. \ref{3}c; see Supplementary Note 15 for moir\'{e} domains with different periods) \cite{zhang_tuning_2018}.
Due to energy level alignment \cite{joshi_control_2014,zimmermann_self-assembly_2020,portner_charge_2020},
this corrugation of $\Phi$  affects substrate-to-MOF electron transfer and hence the effective electron filling of the MOF bands, with this filling smaller at wire than pore regions \cite{joshi_control_2014,kumar_mesoscopic_2023}.
This is consistent with the effective reduction of the hole creation barrier at the wire relative to the pore in Fig. \ref{3}b. 

To capture the effect of this moir\'{e}-induced modulation of $\Phi$ on the MOF electronic properties, we conducted further DMFT calculations. Using $U =$ 0.65 eV (the same as Fig. \ref{1}e), we calculated $A(E)$ for a range of $E_\mathrm{F}$ assuming a uniform system. We considered a sinusoidal variation of $E_\mathrm{F}$ from a minimum value corresponding to half-filling of the kagome MOF bands, with an amplitude of 0.2 eV, to match the experimental $\Delta \Phi$ for this specific hBN/Cu(111) moir\'{e} domain \cite{kumar_mesoscopic_2023,zhang_tuning_2018} (Fig. \ref{3}c; see Methods). The obtained $A(E)$ (Fig. \ref{3}d) reproduce the experimental spectral features in Fig. \ref{3}b, including the shifting of the lower Hubbard band maximum (LHBM) and upper Hubbard band minimum (UHBM) (Fig. \ref{3}f), and the and the vanishing of the gap and increase of the spectral function at the Fermi level for the wire region (Fig. \ref{3}e, g). 

The DMFT-calculated spectral functions $A(E)$ with no gap at the Fermi level at the smallest electron filling ($\Delta \Phi_{\mathrm{DMFT}} = 0.2$ eV) show peaks near $E_\mathrm{F}$ (Fig. \ref{3}d), however, which were not observed in the experimental spectra at the wire region (Fig. \ref{3}b). These peaks are indicative of coherent quasiparticles \cite{georges_dynamical_1996}, with their width associated with the quasiparticle lifetime and quasiparticle mean free path $\ell$. Via our DMFT and TB calculations (Supplementary Note 3), we estimate $\ell \approx 10$ nm, much larger than the wire region width of \ensuremath{\mathord{\sim}}4 nm. We hypothesize that the coherence peaks are suppressed in the experiment as quasiparticles are strongly scattered by the pore regions (where the MOF remains insulating). This is a key difference between Fig. \ref{3}b and d: the experimental measurements represent changes in electron population at finite size MOF regions due to a locally varying work function, whereas each theoretical spectrum corresponds to an infinite uniform system with a constant chemical potential.

Furthermore, some of the experimental spectra in the vicinity of the wire region in Fig. \ref{3}b feature narrow peaks at energies of $\sim$0.4 eV, not present in the theoretical $A(E)$ in Fig. \ref{3}d. We claim that these peaks do not represent intrinsic electronic states, but are instead related to charging phenomena not captured by the DMFT calculations (see Fig. \ref{4} and Supplementary Note 17).

Despite these discrepancies with experiment, our DMFT calculations capture the fundamental electronic properties of the 2D DCA\textsubscript{3}Cu\textsubscript{2} MOF, hosting a Mott insulating phase with $E_{\textrm{g}}\approx 200$ meV, and a metal-like phase (with no gap at the Fermi level) at the wire region (for the specific tip-sample distance considered in Fig. \ref{3}). DMFT is a well-established method for understanding the Mott insulator and Mott MITs \cite{georges_dynamical_1996, kotliar_strongly_2004, vollhardt_dynamical_2012, adler_correlated_2018}. This claim of strong electronic correlations and of a Mott insulating phase for the DCA\textsubscript{3}Cu\textsubscript{2} kagome MOF is consistent with previous literature \cite{kumar_manifestation_2021, fuchs_kagome_2020, field_correlation-induced_2022, ohashi_mott_2006}.

\subsection{Tip-assisted Mott metal-insulator transition}

To explore the nature of the metal-like phase observed at the wire region, we conducted further $\mathrm{d}I/\mathrm{d}V$ STS of the MOF as a function of tip-sample distance $\Delta z + z_0$ (where $z_0$ is set by tunnelling parameters), at a DCA lobe site within the wire region (Fig. \ref{4}d). For large $\Delta z + z_0$, these spectra feature an energy gap $E_{\textrm{g}}$, with a small $\mathrm{d}I/\mathrm{d}V$ signal at the Fermi level (Fig. \ref{4}f), similar to spectra in the pore regions (Figs. \ref{2}a, \ref{3}b; Supplementary Note 20), with a sharp peak at positive $V_{\textrm{b}}$  (purple circles in Fig. \ref{4}d) and a subtler band edge (red squares; similar to band features in Fig. \ref{3}b) at negative $V_{\textrm{b}}$. As $\Delta z$ decreases, the energy position of the sharp peak decreases linearly (Fig. \ref{4}e). Conversely, the energy position of the subtler band edge increases with decreasing $\Delta z$, non-linearly and at a lower rate (Fig. \ref{4}e). These features cross the Fermi level at intermediate $\Delta z$, and the spectrum becomes gapless ($E_{\textrm{g}}= 0$), with a significant increase in $\mathrm{d}I/\mathrm{d}V$ signal at the Fermi level (Fig. \ref{4}f). Note that an intermediate $\Delta z$ was also used for all spectra in Fig. \ref{3}b where similar metal-like signatures were observed at the wire region. We found a similar $\Delta z$-dependent trend for wire region Cu sites (Supplementary Note 21). Notably, moir\'e pore regions remain gapped for all $\Delta z$ values (for both Cu and DCA lobe sites; Supplementary Note 20 and Supplementary Note 21). 

As $V_{\textrm{b}}$ is applied between the tip and Cu substrate, the STM double-barrier tunnel junction (DBTJ) -- where the vacuum between tip and MOF is a first tunnel barrier and the insulating hBN is a second one -- causes a voltage drop at the MOF location. This can lead to energy shifts of MOF states and/or charging of such states when they become resonant with the Cu(111) Fermi level (Fig. \ref{4}a) \cite{kumar_electric_2019,portner_charge_2020,yan_synthesis_2021}. These phenomena have previously been observed for the DCA\textsubscript{3}Cu\textsubscript{2} MOF on a decoupling graphene surface \cite{yan_synthesis_2021}. In this scenario, the bias voltages corresponding to an intrinsic electronic state, $V_\mathrm{state}$, and to charging of such a state, $V_\mathrm{charge}$, vary as a function of $\Delta z$ as:

\begin{equation}\label{DBTJ_state}
 V_\mathrm{state}(\Delta z) = \frac{d_\mathrm{eff}\left(V_{\infty} + \Delta\Phi_{\mathrm{ts}}\right)}{(\Delta z + z_0)} + V_{\infty}, 
\end{equation}

\begin{equation}\label{DBTJ_charge}
 V_\mathrm{charge}(\Delta z) = -\frac{V_{\infty}(\Delta z + z_0)}{d_\mathrm{eff}} - (V_{\infty}+\Delta\Phi_\mathrm{ts}), 
\end{equation}
where $d_\mathrm{eff}$ is the effective width of the hBN tunnel barrier, $V_{\infty}$ is the bias voltage corresponding to the electronic state as $\Delta z \rightarrow \infty$, and $\Delta\Phi_\mathrm{ts}$ is the difference between tip and sample work functions \cite{kumar_electric_2019}.

We fit the $\Delta z$-dependent bias voltage associated with the subtle band edge (red squares) in Fig. \ref{4}d with Eq. \eqref{DBTJ_state}, and the bias voltage associated with the sharp peak (purple circles) with Eq. \eqref{DBTJ_charge} (Fig. \ref{4}e). The agreement between experimental data and fits indicate that the subtle spectral band edge (red) represents an intrinsic MOF electronic state, with its energy shifting as $\Delta z$ varies, and with the sharp peak (purple) corresponding to charging of such a state.

\section{Discussion}

We interpret these results as follows. 
For large $\Delta z$ (Fig. \ref{4}c), the MOF electronic states are strongly pinned to the substrate, and the MOF near-Fermi kagome bands are approximately half-filled. Here, the $\mathrm{d}I/\mathrm{d}V$ spectra feature an energy gap at the Fermi level, both at moir\'e pore (Figs. \ref{2}a, \ref{3}b; Supplementary Note 20 and Supplementary Note 21) and wire regions (bottom spectra of Fig. \ref{4}d; Supplementary Note 21): the entire monolayer 2D MOF is intrinsically a Mott insulator featuring localised electrons. This is consistent with DMFT calculations, which show that the system can remain Mott insulating even with a 0.2 eV modulation of $E_\mathrm{F}$ (corresponding to the experimental moir\'e work function corrugation; Supplementary Fig. 4).

Due to the hBN/Cu(111) moir\'e work function corrugation, the LHBM at a wire region is very close to the Fermi level (in comparison to the pore region). As $\Delta z$ is reduced, the MOF states become less pinned to the Cu(111) substrate and more pinned to the tip via the DBTJ effect (Fig. 4; Supplementary Note 19, Supplementary Note 20, Supplementary Note 21). Given that $\Phi_{\textrm{tip}} > \Phi_{\textrm{wire}}$ (see Supplementary Note 18), this leads to an energy upshift of the MOF states with respect to the Cu(111) Fermi level (with the LHBM susceptible to charging as $V_{\textrm{b}}$ becomes more positive). At intermediate $\Delta z$, this energy upshift depopulates the LHB and leads to the transition from the Mott insulating phase (only existing at half-filling of the three near-Fermi kagome bands; Fig. \ref{1}) to the metallic phase. This is concomitant with a dramatic change in the $\mathrm{d}I/\mathrm{d}V$ spectra, including the vanishing of the energy gap and the increase in $\mathrm{d}I/\mathrm{d}V$ signal at the Fermi level. As $\Delta z$ is further reduced (Fig. \ref{4}a), the near-Fermi kagome bands become fully depopulated (with the bottom of these bands susceptible to charging as $V_{\textrm{b}}$ becomes more negative; top spectra of Fig. \ref{4}d) and the MOF becomes a trivial insulator (with a gap between these near-Fermi bands and lower energy bands; Supplementary Fig. 1). The DBTJ effect also manifests itself at the pore region, with energy shifts of the LHBM and UHBM as $\Delta z$ varies (Supplementary Note 20, Supplementary Note 21). Yet, due to the smaller $\Phi_\mathrm{pore}$ (Fig. \ref{3}c), the Fermi level lies close to the centre of the energy gap (Fig. \ref{3}b), making the Mott insulating phase robust for the considered $\Delta z$ range, consistent with DMFT (Fig. \ref{3}d, Supplementary Figs. 2b, 4).
In the wire region, the STM tip, via the DBTJ effect, in combination with the large $\Phi_{\textrm{wire}}$, acts as a local electrostatic gate, switching the 2D MOF from Mott insulator to metal (Supplementary Fig. 26).

This tip-induced gating at the wire region is inherently local, occurring within the cross section of the DBTJ (typically with a diameter of $\sim$10 nm given by the tip radius of curvature). This locality could lead to $\mathrm{d}I/\mathrm{d}V$ features (e.g., due to electronic confinement) not captured by DMFT. Whether these local changes produce truly delocalised metallic states across an extended area of the sample remains an open question, beyond the scope of this investigation. Future work could investigate the DCA\textsubscript{3}Cu\textsubscript{2} MOF on single-crystal exfoliated hBN, within a gated heterostructure, allowing for uniform electrostatic control and bulk (e.g., transport) measurements.

Our assertion that the MOF is a Mott insulator at the moir\'e pore regions for all tip-sample distances, and at the wire regions for large tip-sample distances, is well supported by DMFT calculations, and by previous literature \cite{kumar_manifestation_2021, fuchs_kagome_2020, field_correlation-induced_2022, ohashi_mott_2006}. The DMFT spectral functions demonstrate excellent agreement with $\mathrm{d}I/\mathrm{d}V$ spectra at the pore regions, including the $\sim$200 meV gap at the Fermi level, and the energy modulation of the LHB and UHB due to variations in electrostatic potential. Our assertion of a metallic phase for the MOF at the wire region for intermediate tip-sample distances is well supported by the DBTJ model, and by the qualitative agreement with DMFT, including the absence of an energy gap at the Fermi level and an increase in Fermi-level $\mathrm{d}I/\mathrm{d}V$ signal and spectral function.

Mott insulating -- with intrinsically localised electronic states -- and metallic phases have been observed in TaS\textsubscript{2} and TaSe\textsubscript{2} monolayers, which feature frustrated lattice geometries and lattice constants (due to charge density wave distortions) similar to our work \cite{lin_scanning_2020,chen_strong_2020,vano_artificial_2021,bu_possible_2019,fei_understanding_2022}, for finite domain sizes as small as $\sim$10x10 nm$^2$ \cite{vano_artificial_2021}. Also, local topological phase transitions induced by an STM tip have been demonstrated \cite{collins_electric-field-tuned_2018}. Our interpretation of a local population-induced Mott metal-insulator transition is consistent with these findings.

Note that the vanishing of the gap at the Fermi level for the metallic phase in the moir\'e wire region me (Fig. \ref{3}b) is reminiscent of spectral features for the pseudogap phase in cuprates and other transition metal oxides \cite{Cai_NatPhys_2016, Battisti_NatPhys_2017, Zhong_PRL_2020}. Future work on controllable Mott MITs in MOFs might contribute to the general understanding of doped Mott insulators.

Monolayer DCA-based MOFs have been studied on other substrates \cite{pawin_surface_2008,yan_synthesis_2021,yan_two-dimensional_2021,hernandez-lopez_searching_2021,kumar_manifestation_2021,kumar_two-dimensional_2018,yan_-surface_2019,lobo-checa_ferromagnetism_2024,vano_emergence_2023}, without observing a Mott phase. In our case, the combination of the wide bandgap hBN as a template (allowing the MOF to retain its intrinsic electronic properties), and of the adequate energy level alignment given by the hBN/Cu(111) substrate (resulting in half-filling of kagome bands; Fig. \ref{1}d), plays a key role in the realisation of the correlated-electron Mott phase.

We have demonstrated that single-layer DCA\textsubscript{3}Cu\textsubscript{2} not only hosts a robust Mott insulating phase (with $E_\textrm{g} \gg k_{\textrm{B}}T$ at $T = 300$ K), but also that Mott MITs can be achieved via the combination of template- (Fig. \ref{3}) and tip- (Fig. \ref{4}) induced gating, consistent with DMFT and the DBTJ model. This shows that such phase transitions can be controlled in monolayer MOFs via electrostatic tuning of the chemical potential.

Our findings represent a promising step towards incorporation of 2D MOFs as active materials in device-like architectures (e.g., van der Waals heterostructures based on 2D materials), benefiting from efficient synthesis approaches and versatility offered by MOFs, and allowing for access and control of correlated-electron phases therein via electrostatic gating \cite{riss_imaging_2014}.
Our work establishes single-layer 2D MOFs -- with crystal geometries allowing for flat bands -- as promising platforms for controllable switching between diverse many-body quantum phenomena, potentially including correlated magnetism, superconductivity, and quantum spin liquids.

\section{Methods}
\subsection{Sample preparation} The monolayer DCA\textsubscript{3}Cu\textsubscript{2} kagome MOF was synthesised on hBN/Cu(111) in UHV (base pressure $\sim$2 $\times 10^{-10}$ mbar). The Cu(111) surface was first cleaned via 2-3 cycles of sputtering with Ar$^+$ ions and subsequent annealing at $\sim$770 K. A hBN monolayer was synthesised on Cu(111) via the thermal decomposition of borazine \cite{auwarter_hexagonal_2019}. We dosed a partial pressure of borazine of ${\sim}9 \times 10^{-7}$ mbar for 45 minutes with the Cu(111) sample maintained at 1140 K. We kept the Cu(111) sample at this temperature for a further 20 minutes to ensure a complete reaction. We then cooled the sample to room temperature and deposited the DCA molecules via sublimation at 390 K, corresponding to a deposition rate of 0.007 ML/sec.  In our experiments we considered DCA coverages of  $\sim$0.4-0.6 ML. We then further cooled the sample to $\sim$77 K before depositing Cu via sublimation at 1250 K (Cu deposition rate: $\sim$0.002 ML/sec; typical Cu coverages in our experiments: $\sim$0.05 ML). Finally, the sample was annealed to $\sim$200 K for 15 minutes. Further details in Supplementary Note 6.

The DCA\textsubscript{3}Cu\textsubscript{2} MOF crystalline structure was found to be commensurate with the hBN lattice but incommensurate with the long-range hBN/Cu(111) moir\'{e} patterns of different sizes, across which the MOF grows without disruption (see Supplementary Fig. 8). The hBN/Cu(111) moir\'{e} pattern is clearly visible in large-scale STM images of the MOF (Fig. \ref{1}a). This is consistent with the modulation of the MOF's electronic properties illustrated in Fig. \ref{3} (also see $\mathrm{d}I/\mathrm{d}V$ maps in Supplementary Fig. 12). 

\subsection{STM and STS measurements} All STM and d$I/$d$V$ STS measurements were performed at 4.5 K (except measurements in Supplementary Note 23 performed at 77 K), at a base pressure ${<}1\times10^{-10}$ mbar, with a hand-cut Pt/Ir tip. All STM images were acquired in constant-current mode with tunnelling parameters as reported in the text (bias voltage applied to sample).  All d$I/$d$V$ spectra were obtained by acquiring $I(V)$ at a constant tip-sample distance (stabilised by a specified setpoint tunnelling current and bias voltage), and by then numerically differentiating $I(V)$ to obtain d$I/$d$V$ as a function of bias voltage. Tips were characterised on regions of bare hBN/Cu(111) prior to spectroscopy measurements, where the Shockley surface state of Cu(111) could be observed (grey curve in Fig. \ref{2}a, onset shifted upwards in energy due to confinement by hBN monolayer) \cite{joshi_boron_2012}.

\subsection{DFT calculations}

We calculated the non-spin-polarised band structure of DCA\textsubscript{3}Cu\textsubscript{2} on hBN on Cu(111) via DFT (Fig. \ref{1}d), using the Vienna Ab-Initio Simulation Package (VASP) \cite{kresse_efficiency_1996} with
the Perdew-Burke-Ernzerhof (PBE) functional under the generalised gradient approximation \cite{perdew_generalized_1996}.
We used projector augmented wave (PAW) pseudopotentials \cite{blochl_projector_1994,kresse_ultrasoft_1999} to describe core electrons, and the semi-empirical potential DFT-D3 \cite{grimme_consistent_2010} to describe van der Waals forces.

The substrate was modelled as a slab three Cu atoms thick, with the bottom layer fixed at the bulk lattice constant \cite{rumble_crc_2019}.
A layer of passivating hydrogen atoms was applied to the bottom face to terminate dangling bonds.

A 400 eV cut-off was used for the plane wave basis set.
The threshold for energy convergence was $10^{-4}$ eV.
The atomic positions of the DCA\textsubscript{3}Cu\textsubscript{2}/hBN/Cu(111)  were relaxed until Hellmann-Feynman forces were less than 0.01 eV/\AA, using a $3\times 3\times 1$ k-point grid for sampling the Brillouin zone and 1st order Methfessel-Paxton smearing of $0.2$ eV.
The charge density for the relaxed structure was calculated using an $11 \times 11 \times 1$ k-point grid, Blochl tetrahedron interpolation, and dipole corrections.
The band structure was determined non-self-consistently from the charge density.

Note that small (${<}1$ \r{A}) perturbations in the height of DCA\textsubscript{3}Cu\textsubscript{2} above the hBN/Cu(111) do not appreciably affect the calculated band structure. As such, small perturbations in height related to the hBN/Cu(111) moir\'{e} pattern (of at most 0.7 \r{A} \cite{schwarz_corrugation_2017}) were not captured by these calculations \cite{field_correlation-induced_2022}.

\subsection{DMFT calculations}

We performed dynamical mean-field theory (DMFT) calculations on the free-standing DCA\textsubscript{3}Cu\textsubscript{2} kagome MOF. We used the Hubbard model for a kagome lattice with nearest-neighbour hopping,
\begin{align}
    H &= -t \sum_{\langle i,j \rangle, \sigma} \hat{c}^\dagger_{i,\sigma} \hat{c}_{j,\sigma} + U \sum_i \hat{n}_{i,\uparrow} \hat{n}_{i,\downarrow},
\end{align}
where the first term is the tight-binding (TB) Hamiltonian with nearest-neighbour hopping energy $t$, $\sum_{\langle i,j \rangle}$ is a sum over nearest-neighbour sites, and the second term is the interaction Hamiltonian with on-site Coulomb repulsion $U$.  The operator $\hat{c}^\dagger_{i,\sigma}$ ($\hat{c}_{i,\sigma}$) creates (annihilates) an electron at site $i$ with spin $\sigma$; $\hat{n}_{i,\sigma} = \hat{c}^\dagger_{i,\sigma} \hat{c}_{i,\sigma}$ is the density operator.
We take $t=0.05$ eV to match prior DFT calculations of DCA\textsubscript{3}Cu\textsubscript{2} \cite{field_correlation-induced_2022,zhang_intrinsic_2016,fuchs_kagome_2020}.

We first calculated the non-interacting ($U=0$) density of states (DOS; blue curve in Fig. \ref{1}e) by numerically integrating over all momenta in the first Brillouin zone. The chemical potential $E_\mathrm{F}$ in Fig. \ref{1}e was chosen to be consistent with the electron filling predicted by DFT (Fig. \ref{1}d). We applied a thermal broadening ($k_\textrm{B} T = 2.5$ meV) to this non-interacting TB DOS to make it consistent with the thermal broadening of the DMFT-generated (see below) spectral function $A(E)$ in Fig. \ref{1}e.

To account for electronic correlations, we then implemented the DMFT formalism \cite{georges_hubbard_1992,georges_dynamical_1996,georges_strongly_2004} using the Toolbox for Researching Interacting Quantum Systems (TRIQS) \cite{parcollet_triqs_2015}, with the continuous-time hybridization expansion solver (CTHYB) \cite{seth_triqscthyb_2016,gull_continuous-time_2011} at a temperature of $\sim$29 K ($k_{\textrm{B}}T \approx 0.05t$; unless specified otherwise, see Supplementary Note 23 for temperature-dependent calculations), with $U = 0.65$ eV.
To use a single-site DMFT formalism \cite{georges_dynamical_1996} with the kagome band structure, the non-interacting DOS of the three kagome bands were combined into a single function for use as the input into the DMFT procedure.

We calculated the many-body spectral functions $A(E)$ (analogous to the DOS, but in the interacting regime; Fig. \ref{3}d) via analytic continuation using the maximum entropy method (MaxEnt) as implemented in TRIQS \cite{kraberger_maximum_2017}.
The meta-parameter, $\alpha$, was determined from the maximum curvature of the distance between the MaxEnt fit and data, $\chi^2$, as a function of $\alpha$ \cite{bergeron_algorithms_2016}.

Each DMFT calculation assumed a spatially uniform work function $\Phi$; long-range modulation of $\Phi$ is beyond the capabilities of DMFT.
As such, the spatially varying sample work function $\Phi$ resulting from the experimental hBN/Cu(111) moir\'e pattern was not explicitly captured in the individual $A(E)$ spectra in Fig. \ref{3}d.
This spatial variation of $\Phi$ was approximated by varying the uniform $E_\mathrm{F}$ of the system for each individual $A(E)$ spectrum. Each of these $A(E)$ spectra was then associated to a specific location of the hBN/Cu(111) moir\'e pattern (and hence to a specific experimental $\textrm{d}I/\textrm{d}V$ curve) based on how this variation of $E_\mathrm{F}$ would translate to a local $\Phi$.
These calculations assume that the theoretical $A(E)$ spectra, calculated with a uniform $E_F$ (and hence uniform $\Phi$), are reasonable representations of the locally acquired experimental d$I$/d$V$ curves, which are affected by a spatially varying $\Phi$. This assumption is reasonable for the insulating phase. Indeed, in the Mott insulating phase, electronic states are localised at the kagome sites, confined within areas that are small in length \cite{fazekas_lecture_electron_correlation_1999} compared to the distance between nearest-neighbour kagome sites ($\sim$1 nm) and to the periodicity $\lambda$ of the hBN/Cu(111) moir\'e domains considered in our experiments ($\lambda > 5$ nm). DMFT indicates that this Mott insulating phase is robust to variations in chemical potential $E_\textrm{F}$ larger than 0.2 eV, with the spectral function $A(E)$ shifting in energy as $E_\textrm{F}$ is varied within this range,  without other significant qualitative changes (see Supplementary Fig. 2b for $E_\textrm{F}=$ 0.25 to 0.5 eV). In our experiments, the hBN/Cu(111) moir\'e pattern imposes a periodic modulation of the local work function $\Phi$, with a peak-to-peak modulation amplitude of $\sim$0.2 eV and a modulation periodicity $\lambda \approx 12.5$ nm (Fig. \ref{3}c; this amplitude becomes smaller with decreasing $\lambda$, see Supplementary Fig. 22). That is, $\Phi$ varies \emph{slowly} across the MOF kagome lattice. The effect of such long-range modulation of $\Phi$ on the MOF localised electronic states is to shift the energy of these localised states accordingly. As long as the $\Phi$ modulation period is larger than the distance between nearest-neighbour kagome sites and the $\Phi$ modulation amplitude is smaller than a critical value inducing the transition to the metallic phase, there is no other dramatic qualitative effect on these localised electronic states. This explains the excellent agreement between experimental $\mathrm{d}I/\mathrm{d}V$ spectra for the MOF at the hBN/Cu(111) moir\'e pore region and DMFT-calculated spectral functions $A(E)$ for the system in the Mott insulating phase (Fig. \ref{3}b, d), with LHB and UHB modulated in energy following the variation in electrostatic potential. For the MOF in the metallic phase at the moir\'e wire region (Figs. \ref{3}, \ref{4}), discrepancies between theory and experiment can be explained (as discussed in the main text) by long electronic coherence lengths (Supplementary Note 3), and by effects of long-range moir\'e $\Phi$ modulation and of finite DBTJ cross section on the potentially delocalised metallic MOF states.

\subsection{DBTJ model}
In Figs. \ref{4}d, e, the bias voltage, $V_{\textrm{state}}$, corresponds to the energy level of an intrinsic MOF frontier electronic state (which is susceptible to charging), at a band edge in either the Mott or trivial insulator regime. Red square markers indicating $V_{\textrm{state}}$ were assigned based on method outlined in Supplementary Note 13. The bias voltage, $V_\mathrm{charge}$, corresponds to the peak associated with charging of such a MOF state. Purple circle markers indicating $V_\mathrm{charge}$ were assigned by finding a local maximum in $\mathrm{d}I/\mathrm{d}V$.

In Fig. \ref{4}e, we considered four experimental datasets for fitting with Eqs. \eqref{DBTJ_state} and \eqref{DBTJ_charge}: $V_{\textrm{state}}(\Delta z)$ (red squares) and $V_{\textrm{charge}}(\Delta z)$ (purple circles) for small values of $\Delta z$ (trivial insulator phase), and $V_{\textrm{state}}(\Delta z)$ and $V_{\textrm{charge}}(\Delta z)$ for large values of $\Delta z$ (Mott insulator phase). Given the Mott MIT, we considered two different intrinsic MOF band edges susceptible of charging, embodied in two different values of $V_{\infty}$: one for the trivial insulator phase (small values of $\Delta z$) and one for the Mott insulator phase (large $\Delta z$). This phase transition is evident from the offset in $V_{\textrm{state}}$ (red squares in Fig. \ref{4}e) observed when $\Delta z$ varies from small to large (through the metallic phase at intermediate $\Delta z$). Accordingly, we used a global fitting approach to obtain the same fitting parameters $d_\mathrm{eff}$, $z_0$, and $\Delta\Phi_\mathrm{ts}$ (characteristic of the DBTJ and the acquisition location) for these four experimental datasets, and a separate $V_{\infty}$ value for each regime.

Schematics in Fig. \ref{4}a-c represent cartoon illustrations of our proposed physical mechanism for tip-induced gating.

It is important to note that the DBTJ inherently affects all $\mathrm{d}I/\mathrm{d}V$ measurements in this work, including those in Figs. \ref{2} and \ref{3}, for both pore and wire regions of the hBN/Cu(111) moir\'e pattern. The DBTJ effect does not cause phase transitions at the pore regions, however (see Supplementary Note 20, Supplementary Note 21). The measurements in Fig. \ref{3} were performed with an intermediate tip-sample distance -- which is why metallic properties were observed at the wire region.

\section{Data availability}

The data supporting the findings of this study is available from the authors upon request.

\section{Code availability}

All codes relating to DMFT and MaxEnt are available at \url{https://doi.org/10.5281/zenodo.7439858}. Code for data analysis and theoretical calculations can be made available from the authors upon request.

\bibliography{hBN_MOF_paper.bib}

\section{Acknowledgements}
A.S. acknowledges funding support from the ARC Future Fellowship scheme (FT150100426). B.J.P. acknowledges funding support from the ARC Discovery Project scheme (DP180101483). H.L.N. acknowledges funding support from the MEXT Quantum Leap Flagship Program (JPMXS0118069605). B.L., J.H., J.C., and N.V.M. acknowledge funding support from the Australian Research Council (ARC) Centre of Excellence in Future Low-Energy Electronics Technologies (CE170100039). B.L., B.F., and J.C. are supported through Australian Government Research Training Program (RTP) Scholarships. B.F. and N.V.M.  gratefully  acknowledge  the  computational  support  from  National  Computing Infrastructure and Pawsey Supercomputing Facility. The authors also thank Prof. Michael S. Fuhrer, Prof. Jaime Merino Troncoso, and Dr. Daniel Moreno Cerrada for valuable discussions. 

\section{Author Contributions}
B.L., J.H., and A.S. conceived and designed the experiments. B.L., J.H., and J.C. performed the experiments. B.F. and H.L.N. performed the theoretical calculations with guidance from B.J.P. and N.V.M. All authors contributed to writing the manuscript. 

\section{Competing Interests}
The authors declare no competing interests.

\section{Additional Information}
\subsection{Supplementary Information}
Further DFT calculations; further DMFT calculations; discussion of quasiparticle peaks in Fig. 3; MOF growth and structure details; comparison between DMFT and STS; bias-dependent STM topography images and $\mathrm{d}I/\mathrm{d}V$ maps; further STS measurements; STS measurements across moir\'e patterns with different periods; method for determining band edges; further charging signatures; estimate of tip work function; further DBTJ model details including tip-induced gating at Cu sites of the MOF; $\mathrm{d}I/\mathrm{d}V(\Delta z)$ measurements at the pore region; further evidence of metallic character; temperature-dependent STS measurements and DMFT calculations; comparison of STS measurements at different moir\'e pore regions. 

\subsection{Correspondence \& Requests for Materials}
To be addressed to Ben J. Powell, Nikhil V. Medhekar, or Agustin Schiffrin.

\begin{figure*}[h]
    \centering
    \includegraphics[width=\linewidth]{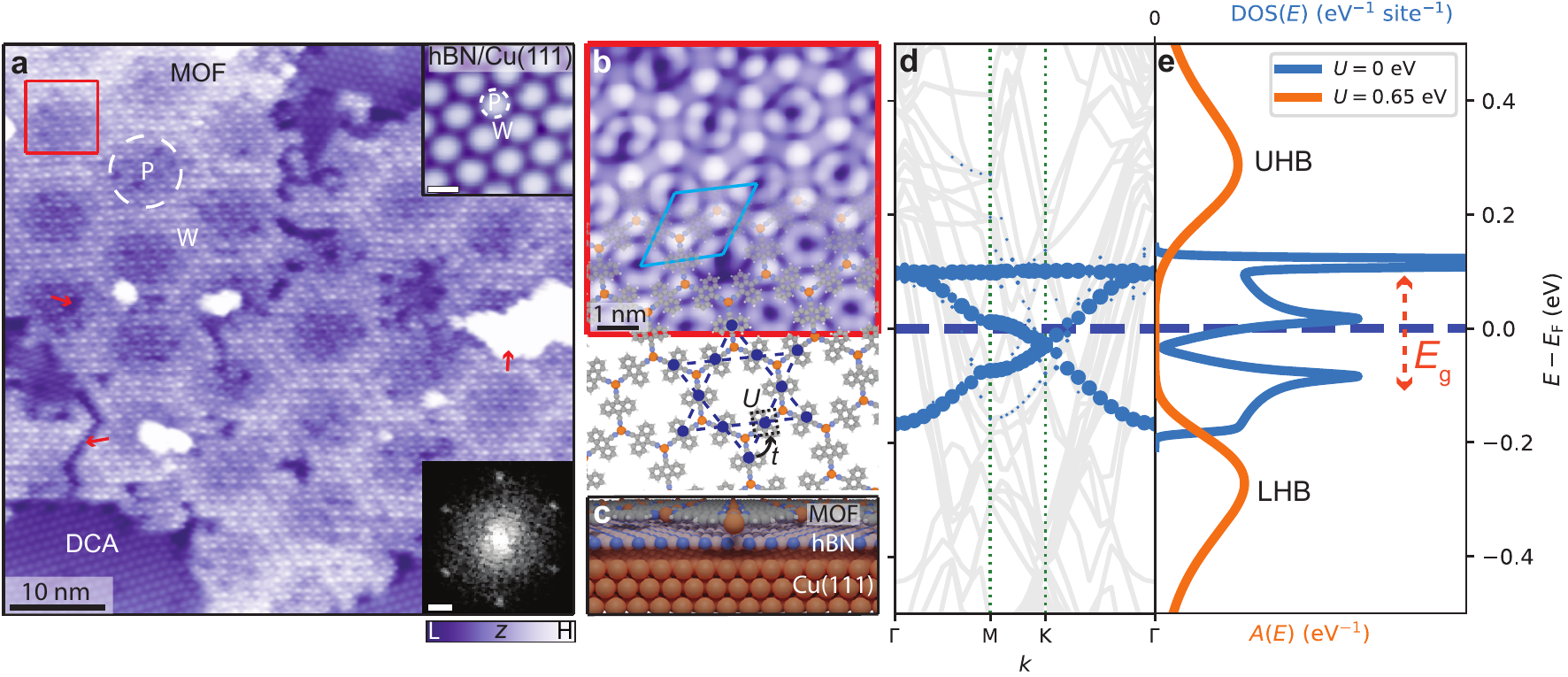}
    \caption{\textbf{A 2D kagome MOF on an atomically thin insulator: DCA\textsubscript{3}Cu\textsubscript{2} on single-layer hBN on Cu(111).
    a,} STM image of MOF (with organic DCA-only regions; $V_\textrm{b}$ = -1 V, $I_\textrm{t}$ = 10 pA).
    `P' and `W' indicate pore (dashed white circle) and wire regions of hBN/Cu(111) moir\'{e} pattern.
    Lower inset: Fourier transform of STM image; sharp spots correspond to MOF hexagonal periodicity (scale bar: 0.25 nm$^{-1}$). Upper inset: STM image of bare hBN/Cu(111) moir\'{e} pattern ($V_\textrm{b}$ = 4 V, $I_\textrm{t}$ = 100 pA, scale bar: 4 nm). Red arrows: examples of crack, vacancy, and Cu cluster defects within MOF domain.
    \textbf{b,} Top-view of MOF model overlaid upon small-scale STM image of region within red box in (a) ($V_\textrm{b}$ = -1 V, $I_\textrm{t}$ = 10 pA).
    MOF unit cell indicated in light blue. Blue dashed lines and solid circles: kagome pattern formed by DCA molecules, with inter-site electron hopping, $t$, and on-site Coulomb repulsion, $U$. \textbf{c,} Model of MOF/hBN/Cu(111) (side view). Hydrogen: white; carbon: grey; boron: pink; nitrogen: blue; copper: orange.
    \textbf{d,} Electronic band structure calculated by DFT (with $U = 0$) \cite{field_correlation-induced_2022}. Blue circles: projections onto MOF states. Grey curves: Cu(111) states (hBN states do not contribute within the shown energy range).
    Hybridisation between MOF and Cu(111) is hindered by the hBN monolayer; the MOF band structure retains its kagome character. Chemical potential $E_{\textrm{F}}$ (blue dashed line) is close to half-filling of kagome bands.
    \textbf{e,} Density of states, DOS($E$) (tight-binding model with thermal broadening, $U$ = 0, blue), and spectral function, $A(E)$ (DMFT, $U =$ 0.65 eV, orange), of free-standing MOF. Local Coulomb interaction opens a significant Mott energy gap $E_{\textrm{g}}$ between lower (LHB) and upper Hubbard bands (UHB). }
    \label{1}
\end{figure*}

\begin{figure*}[h]
    \centering
    \includegraphics[width=\linewidth]{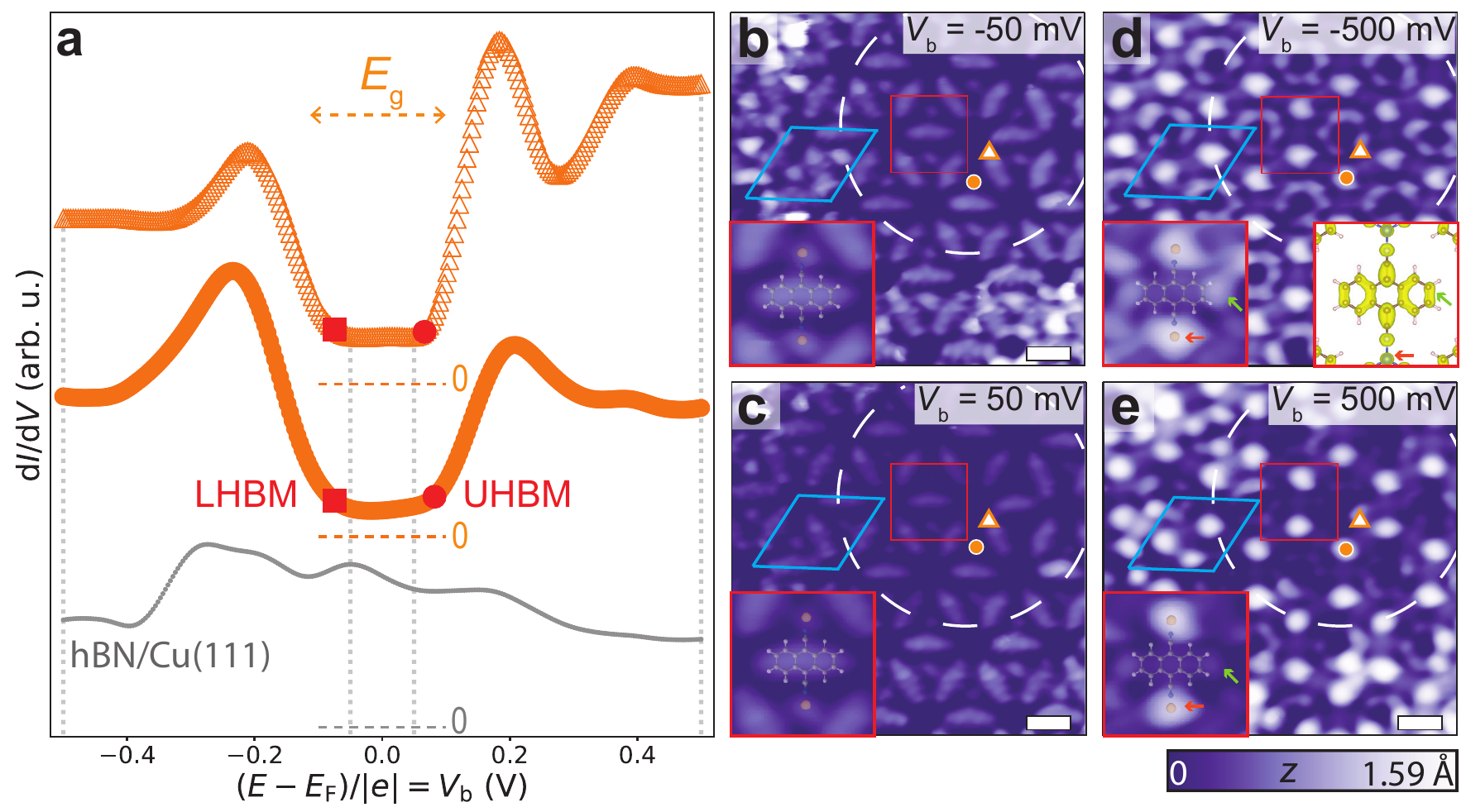}
    \caption{\textbf{Bandgap \textbf{\textit{E}}$_{\textbf{g}}$ $\mathbf{\approx 200}$ meV in DCA\textsubscript{3}Cu\textsubscript{2} MOF on hBN/Cu(111).
    a,} $\textrm{d}I/\textrm{d}V$ spectra at MOF positions indicated by orange markers in (b-e) and on bare hBN/Cu(111) (grey dots) (setpoint: $V_\textrm{b}$ = -500 mV, $I_\textrm{t}$ = 500 pA).
    Gray dotted vertical lines indicate bias voltages at which STM images were acquired in (b-e).
    Spectra offset for clarity. Dashed horizontal lines indicate $\mathrm{d}I/\mathrm{d}V = 0$ reference for each curve.
    Spectra reveal a bandgap $E_{\textrm{g}} \approx 200$ meV. The non-zero $\textrm{d}I/\textrm{d}V$ within gap is due to states of underlying Cu(111) leaking through hBN.
    Red squares (circles): lower Hubbard band maxima, LHBM (upper Hubbard band minima, UHBM).
    \textbf{b-e,} STM images of MOF on hBN/Cu(111) at specified bias voltages ($I_\textrm{t}$ = 10 pA). White dashed circle: hBN/Cu(111) moir\'{e} pore. MOF unit cell indicated in light blue. Scale bars: 1 nm. Insets: zoom-in of region within red box, with overlaid Cu-DCA-Cu chemical structure, showing significant contributions from the ends of the DCA anthracene moiety (green arrow) and Cu (orange arrow) for $V_\textrm{b} < \mathrm{LHBM}$ and $V_\textrm{b} > \mathrm{UHBM}$. Right inset in (d): charge density isosurface (0.0025 e$^-$\r{A}$^{-3}$) of DCA\textsubscript{3}Cu\textsubscript{2} obtained by integration of near-Fermi ($\pm 0.5$ eV) DFT wavefunctions.}
    \label{2}
\end{figure*}

\begin{figure*}[h]
    \centering
    \includegraphics[width=\linewidth]{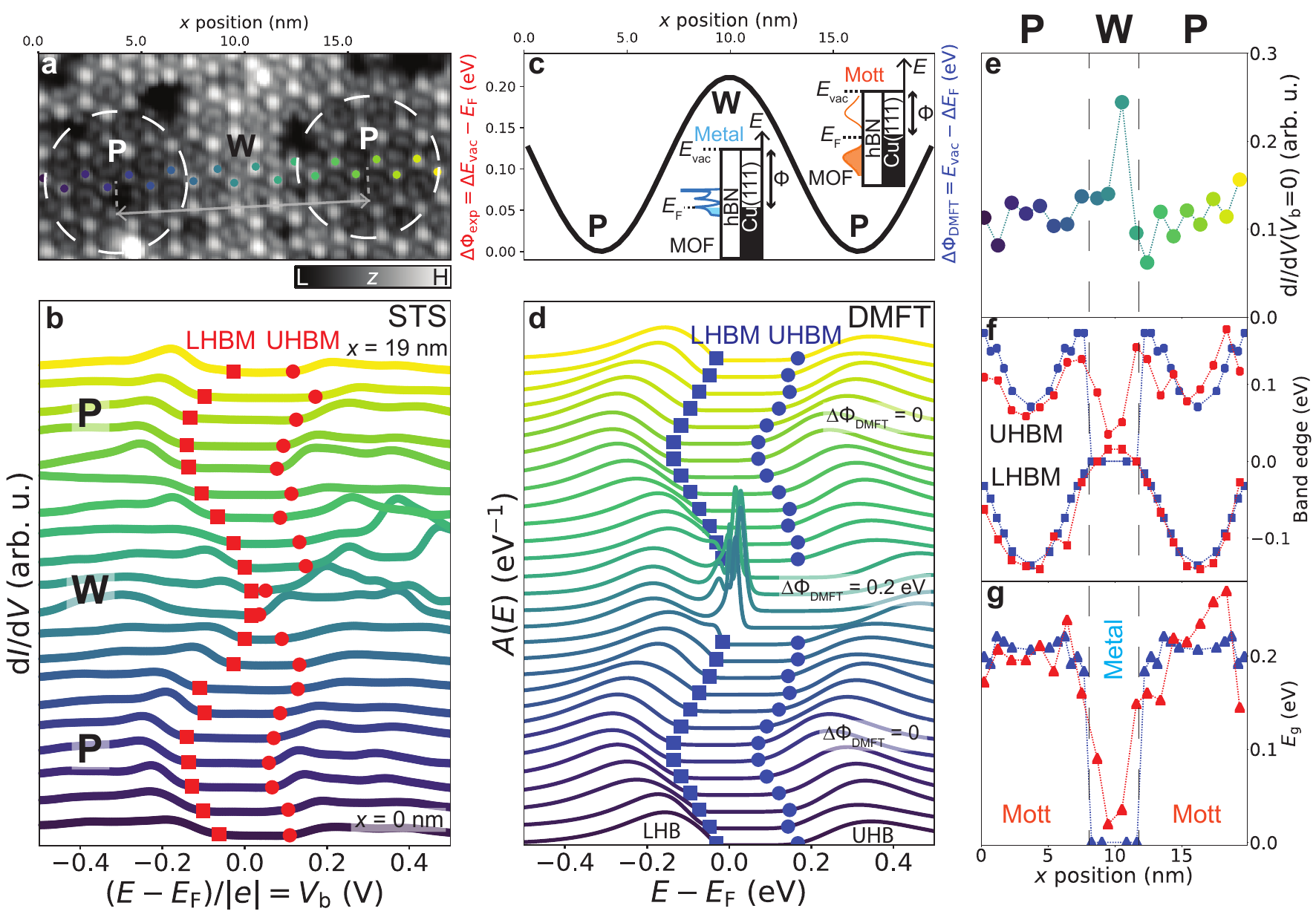}
    \caption{\textbf{Variation of Mott energy gap in DCA\textsubscript{3}Cu\textsubscript{2} MOF induced by hBN/Cu(111) moir\'{e} modulation of local work function. a,} STM image of MOF ($V_\textrm{b}$ = -1 V, $I_\textrm{t}$ = 10 pA). White dashed circles (P): hBN/Cu(111) moir\'{e} pores, separated by wire (W). Grey arrow indicates moir\'{e} period $\lambda \approx$ 12.5 nm.
    \textbf{b,} d$I/$d$V$ spectra acquired at MOF Cu sites, at positions indicated by coloured markers in (a) (tip 190 pm further from STM setpoint $V_\textrm{b}$ = 10 mV, $I_\textrm{t}$ = 10 pA). Energy gap $E_{\textrm{g}}\approx200$ meV at P regions, vanishing at W region (LHBM: lower Hubbard band maximum; UHBM: upper Hubbard band minimum).
    \textbf{c,} Sinusoidal variation of work function, $\Delta \Phi=\Delta E_{\textrm{vac}}-E_{\textrm{F}}$ ($E_{\textrm{vac}}$: vacuum energy level), across hBN/Cu(111) moir\'{e} domain with periodicity $\lambda \approx$ 12.5 nm \cite{zhang_tuning_2018}, affecting the MOF electron filling.
    \textbf{d,} Spectral functions $A(E)$ calculated via DMFT ($U=0.65$ eV, $t=0.05$ eV) for isolated uniform DCA\textsubscript{3}Cu\textsubscript{2}, for different values of $E_{\textrm{F}}$. We account for experimental corrugation $\Delta\Phi$ by varying $E_{\textrm{F}}$ sinusoidally with an amplitude of 0.2 eV as per (c) (see Methods).
    \textbf{e,} Experimental $\mathrm{d}I/\mathrm{d}V$ signal at Fermi level ($V_\mathrm{b} = 0$) as a function of $x$ position, from (b). Increased $\mathrm{d}I/\mathrm{d}V(V_\mathrm{b} = 0$) indicates metallic phase.
    \textbf{f, g,} Experimental (red; from b) and DMFT (blue; from d) LHBM (squares), UHBM (circles) and energy gap $E_{\textrm{g}}$ (triangles), as a function of \textit{x} in (a) (experiment) or corresponding $E_{\textrm{F}}$ (DMFT).
    }
    \label{3}
\end{figure*}

\begin{figure*}[h]
    \centering
    \includegraphics[width=\linewidth]{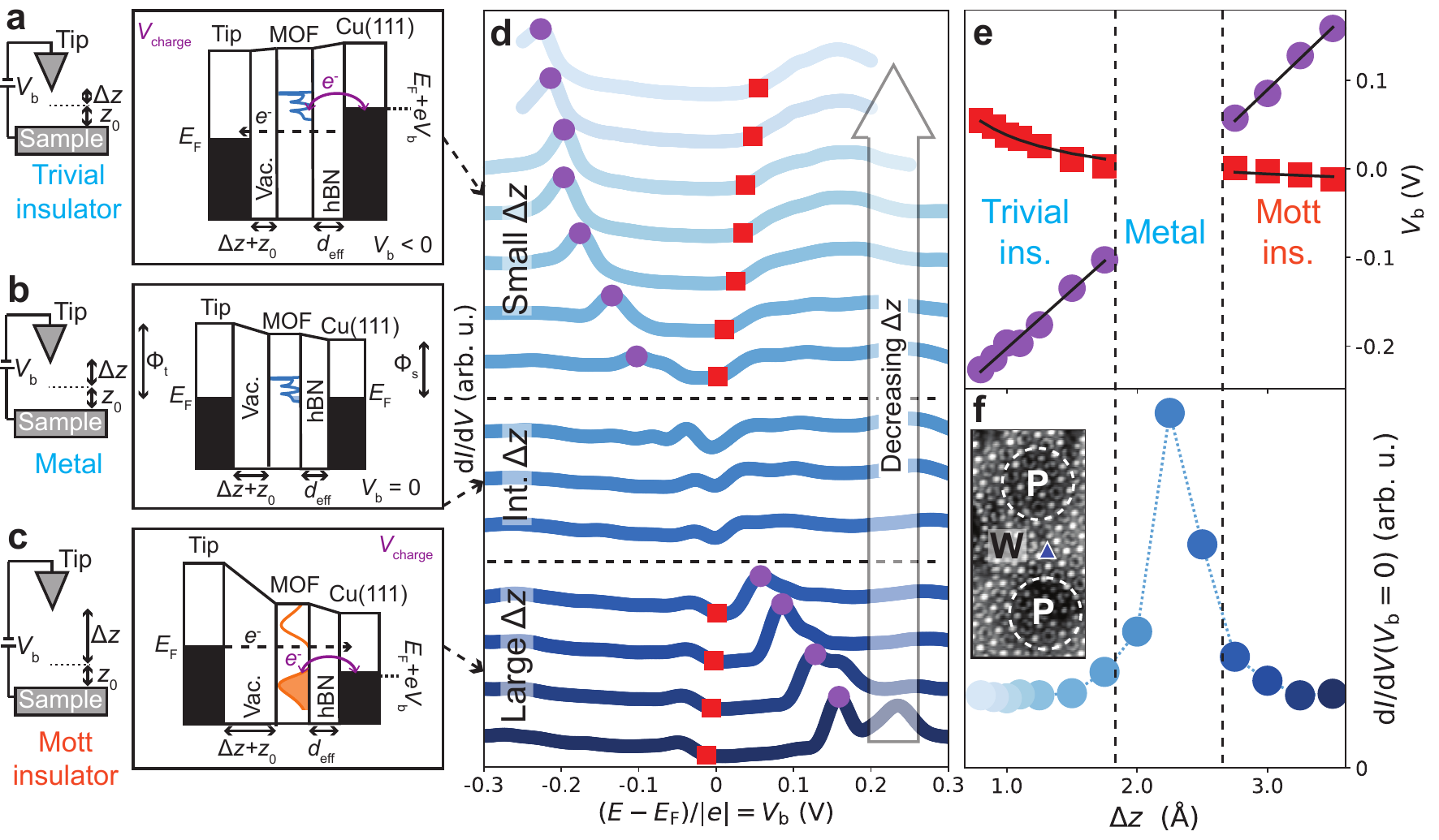}
    \caption{\textbf{Mott insulator-to-metal transition controlled via tip-induced gating.} \textbf{a-c,} Schematics and energy diagrams of tunnelling and charging processes at a double-barrier tunnel junction (DBTJ), consisting of STM tip, vacuum barrier (vac.), MOF, hBN barrier and Cu(111), for small, intermediate (int.) and large tip-sample distances ($\Delta z + z_\mathrm{0}$). When $V_\textrm{b}$ is applied, voltage drop at MOF location enables tip-controlled charging, energy level shifts and gating of MOF transitions, from (correlated) Mott insulator to metal to trivial insulator. These schematics qualitatively illustrate our proposed physical mechanism of tip-induced gating.
    \textbf{d,} d$I$/d$V$ spectra at MOF DCA lobe site, at hBN/Cu(111) moir\'{e} wire, for different $\Delta z + z_\mathrm{0}$ ($z_\mathrm{0}$ given by STM setpoint $V_{\textrm{b}}=$ 10 mV, $I_{\textrm{t}}=$ 10 pA). Purple circles (red squares): MOF charging peak (intrinsic electronic state at MOF band edge). Spectra normalised and offset for clarity.
    \textbf{e,} $V_{\textrm{charge}}$ (purple circles in d) and $V_{\textrm{state}}$ (red squares in d) as a function of $\Delta z$. Black solid lines: global fits to Eqs. \eqref{DBTJ_state} and \eqref{DBTJ_charge}.  
    \textbf{f,} $\mathrm{d}I/\mathrm{d}V$ signal at Fermi level ($V_\mathrm{b} = 0$) as a function of $\Delta z$, from (d). Increased $\mathrm{d}I/\mathrm{d}V(V_\mathrm{b} = 0$) indicates metallic phase at intermediate $\Delta z + z_0$. 
    Inset: STM image of MOF on hBN/Cu(111) showing site (blue triangle marker) where $\mathrm{d}I/\mathrm{d}V(\Delta z)$ measurements were performed ($V_{\mathrm{b}} = -1$ V, $I_{\mathrm{t}} = 10$ pA, scale bar: 2 nm).}
    \label{4}
\end{figure*}
\clearpage


\section*{Supplementary material}

# Supplementary Note 1 DFT calculations: large energy-window band structure

In the main text, we showed DFT calculations of the $DCA_3Cu_2$/hBN/Cu(111) band structure over an energy window from -0.5 to 0.5 eV (Fig. 1d). Supplementary Fig. 1 shows the same calculated band structure over a wider energy window from -3 to 3 eV, for both the freestanding MOF (Supplementary Fig. 1a) and the MOF on hBN/Cu(111) (Supplementary Fig. 1b). These calculations show that the three near-Fermi MOF kagome bands are well isolated and separated in energy by more than 1 eV from the other bands. This is why we focus on these near-Fermi electronic states throughout the main text.

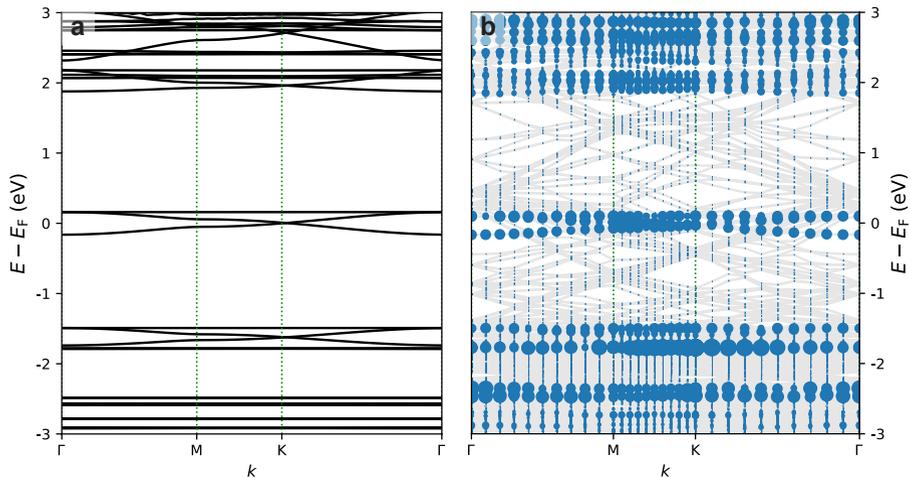

**Supplementary Fig. 1**: **DFT band structures of $DCA_3Cu_2$ MOF over wide energy range. a,** Freestanding MOF. **b,** MOF on hBN/Cu(111). Grey curves: Cu(111) and hBN bands. Blue circles: projections onto MOF states.

# Supplementary Note 2 Dynamical mean-field theory calculations

We performed dynamical mean-field theory (DMFT) calculations for the freestanding kagome lattice (see Methods in main text), for different values of

Hubbard on-site Coulomb repulsion $U$ and chemical potential $E_F$. The chemical potential is referenced with respect to the Dirac point of the free-standing kagome lattice density of states (DOS) (i.e., the $U = 0$ spectral function $A(E)$). We used a nearest-neighbour hopping energy $t = 0.05$ eV, matching the DFT-calculated band structure [1]. We calculated $A(E)$ for different values of $U$ at half-filling (i.e., one electron per site, with $E_F \approx 3t/2 + U/2$), and for different values of $E_F$ at a fixed $U = 0.65$ eV, see Supplementary Fig. 2.

Supplementary Fig. 2a shows a Mott insulating phase emerging at half-filling for all $U \geq U_{crit} \approx 0.55$ eV. The energy gap increases with increasing $U$ (see dashed lines in Supplementary Fig. 2a). For $U < U_{crit}$, the on-site Coulomb repulsion is not sufficiently large (relative to the bandwidth 0.3 eV of the non-interacting kagome band structure) to open a Mott gap; the spectral functions $A(E)$ resemble the non-interacting (i.e., single-electron) DOS.

Supplementary Fig. 2b displays the variation of $A(E)$ with $E_F$, at a fixed $U = 0.65$ eV. At this value of $U$, the bandgap and Mott insulating phase persist across a chemical potential window of $\sim 0.25$ eV ($5t \leq E_F \leq 9.5t$). When $E_F$ increases (decreases), these bands progressively shift to lower (higher, respectively) energy. Outside this range of $E_F$, the Mott phase breaks down and $A(E)$ starts resembling again the non-interacting DOS given by the tight-binding model (when $U = 0$).

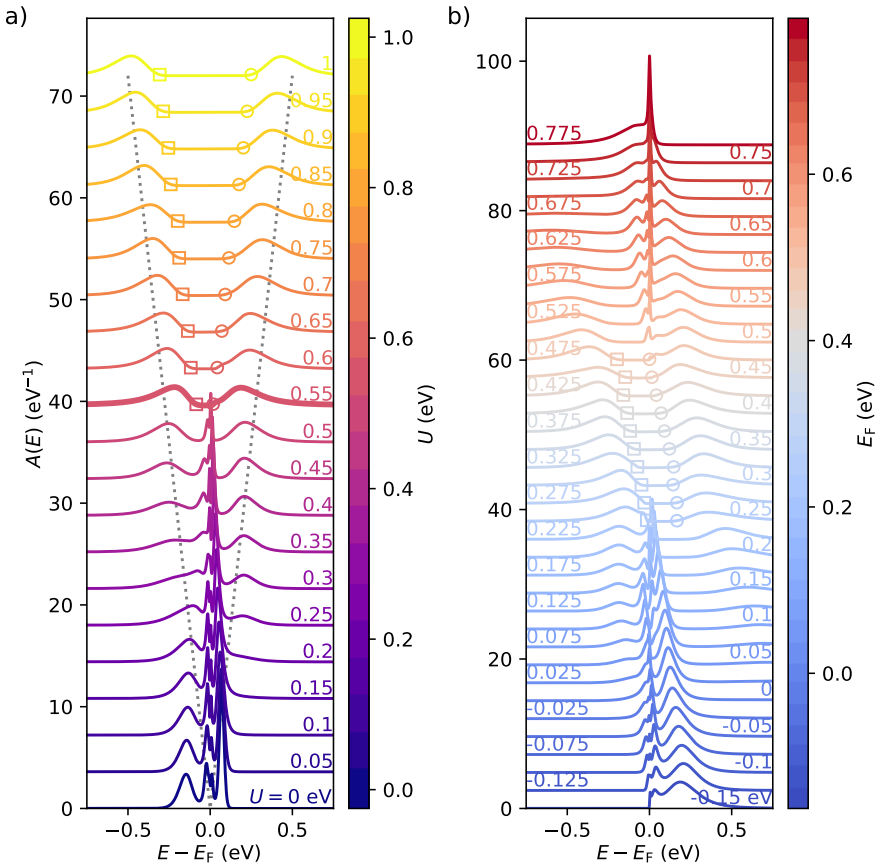

**Supplementary Fig. 2**: **Spectral functions $A(E)$ calculated by DMFT for the free-standing kagome lattice. a,** Spectra for different on-site Coulomb repulsion energies, $U$, with chemical potential, $E_F$ adjusted to maintain half-filling of the kagome three-band band structure ($E_F \approx 1.5t + U/2$; $t = 0.05$ eV). The spectra for $U < U_{crit}$, where $U_{crit} \approx 0.55$ eV (bold spectrum), correspond to the trivial metallic phase. For $U > U_{crit}$, the spectra reveal a correlated-electron Mott insulating phase. Grey dashed lines mark $\pm U/2$, where the Hubbard bands are expected to be observed. Squares (circles): lower Hubbard band maxima, LHBM (upper Hubbard band minima, UHBM, respectively). **b,** Spectral functions for different values of $E_F$, at fixed $U = 0.65$ eV. Spurious dips at $E = E_F$ and additional broadening away from $E_F$ are due to limitations of analytic continuation approach [2].

The spectral functions $A(E)$ in Supplementary Fig. 2b correspond to those in Fig. 3d of the main text. In Fig. 3c, d of the main text, we determined the correspondence between experimental d$I$/d$V$ spectra – acquired at different

positions $x$ along two pore regions of the hBN/Cu(111) moiré domain in Fig. 3a of the main text – and theoretical $A(E)$ curves (each calculated for a uniform system and characterised by a specific chemical potential $E_F$), by assuming a sinusoidal variation of the experimental local work function, $\Delta\Phi$, as a function of $x$. For a periodicity $\lambda \approx 12.5$ nm for this specific moiré domain, $0.15 \lesssim \Delta\Phi \lesssim 0.3$ eV based on experimental estimates [3]. In our calculations, we took account of this experimental $\Delta\Phi(x)$ by varying the chemical potential $E_F$ of the uniform model kagome system accordingly. A chemical potential variation amplitude $\Delta E_F = 0.2$ eV (Fig. 3c of main text) yielded, for the DMFT-calculated $A(E)$ curves, lower Hubbard band maxima and upper Hubbard band minima that matched simultaneously those of our experimental d$I$/d$V$ results (Fig. 3e of main text).

# Supplementary Note 3  Absence of quasiparticle peaks in metallic regime

Some of the DMFT-calculated spectral functions $A(E)$ in Fig. 3d of the main text and Supplementary Fig. 2 show sharp peaks near $E = E_F$ in the metallic regime. We attribute these peaks to Landau Fermi-liquid quasiparticles [4]. These peaks are not observed in the experimental $dI/dV$ spectra in Fig. 3b of the main text. In our DMFT calculations the peaks correspond to quasiparticles with a mean free path $\ell \gtrsim 10$ nm. This is of the order of the period of the hBN/Cu(111) moiré domain, and is much larger than the width of the hBN/Cu(111) moiré wire regions ($\sim 4$ nm). Hence, since quasiparticles can only propagate coherently over the finite width of the wire regions and are strongly scattered by the pore regions, the quasiparticle peak is suppressed.

In what follows we rationalise our estimate of $\ell \gtrsim 10$ nm. The mean-free path of an electron near the Fermi surface is given by $\ell_{\mathbf{k}} = -\hbar v(\mathbf{k})/(2\mathrm{Im}[\Sigma(\mathbf{k}, E)])$ [5], where $v(\mathbf{k})$ is the magnitude of the non-interacting group velocity, and $\Sigma(\mathbf{k}, E)$ is the many-body self-energy. In a Fermi liquid, as $T \to 0$, $\mathrm{Im}[\Sigma(\mathbf{k}, E)] \to 0$ at the Fermi surface ($E \to E_F$ and $\mathbf{k} \to \mathbf{k}_F$) and the mean-free path is infinite. In DMFT, $\Sigma(\mathbf{k}, E)$ is approximated to have no dependence on $\mathbf{k}$; we can hence make the replacement $\Sigma(\mathbf{k}, E) \to \Sigma(E)$.

For the two bottom Dirac bands of the band structure of the non-interacting kagome system, the magnitude of the group velocity is

$$v(\mathbf{k}) = \frac{at}{\hbar} \sqrt{\frac{\sin^2\left(\frac{ak_x}{2}\right)\left(2\cos\left(\frac{ak_x}{2}\right) + \cos\left(\frac{\sqrt{3}ak_y}{2}\right)\right)^2 + 3\cos^2\left(\frac{ak_x}{2}\right)\sin^2\left(\frac{\sqrt{3}ak_y}{2}\right)}{1 + 4\cos^2(ak_x/2) + 4\cos(ak_x/2)\cos(\sqrt{3}ak_y/2)}}.$$

(1)

where $a$ is the kagome lattice constant. The non-interacting Fermi surface is given by the wavevectors satisfying

$$k_y = \pm\frac{2}{\sqrt{3}a}\arccos\left(\frac{(E_F/2t)^2 - 1/4 - \cos^2(ak_x/2)}{\cos(ak_x/2)}\right).$$

(2)

For a value of $E_F$ corresponding to an electron occupation [6] of $n = 0.915$ in the middle of the wire region ($n$ taken from the calculation in Supplementary Fig. 3, with $E_F$ differing here due to using $U = 0$ for calculating $v_F$; $n = 2$ corresponds to fully occupied kagome bands), using $a = 2.0432$ nm, the average non-interacting Fermi velocity is $v_F = (1.3 \pm 0.3) \times 10^5$ ms$^{-1}$, where $\pm$ indicates the range of values $v(\mathbf{k})$ takes at $E = E_F$. Note that at $n = 1$ (half-filling), $v_F = 1.4 \times 10^5$ ms$^{-1}$.

In our DMFT calculations, $\Sigma(E)$ is computed as a function of discrete Matsubara frequencies [5], $\omega_n = (2n + 1)\pi k_B T/\hbar$, where $n$ is an integer, $k_B$ is the Boltzmann constant, and $T$ is the temperature. In order to calculate $\Sigma(E)$ (for arbitrary energy $E$), we can perform analytic continuation [2] to real frequencies, $\omega = E/\hbar$. This procedure is computationally costly due to noise

in numerical quantum Monte Carlo calculations. However, $\Sigma(E = E_F \pm \pi k_B T)$ can be obtained from the lowest Matsubara frequency by the straightforward analytic continuation, $i\omega_n \rightarrow \omega - E_F\hbar + i\eta$, where $\eta$ is an infinitesimal real number. Since $\text{Im}[\Sigma(i\omega_n)]$ is a monotonic function in the vicinity of $\omega_{n=0}$, $|\text{Im}[\Sigma(i\omega_{n=0})]|$ is an overestimate of $|\text{Im}[\Sigma(E = E_F)]|$ by analytic continuation. Thus, $\text{Im}[\Sigma(i\omega_{n=0})]$ provides an underestimate of the mean-free path of a typical quasiparticle, as statistically most occupied states have an energy $\omega < \omega_{n=0}$. Therefore, the mean-free path of a typical quasiparticle can be estimated as

$$\ell \gtrsim -\frac{\hbar v_F}{2\text{Im}[\Sigma(i\omega_{n=0})]}. \tag{3}$$

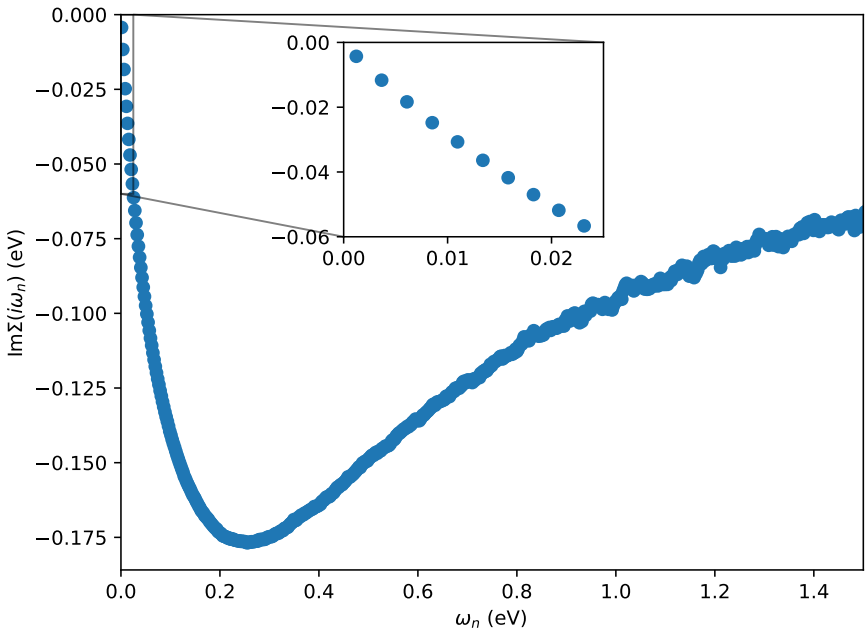

**Supplementary Fig. 3**: Imaginary part of the self-energy $\Sigma(i\omega_n)$ as a function of Matsubara frequencies $\omega_n$, calculated by DMFT for $U = 0.65$ eV, $E_F = 0.2$ eV and $T = 4.5$ K.

Figure 3d of the main text shows spectral functions $A(E)$ calculated via DMFT, with $U = 0.65$ eV, for different chemical potentials $E_F$ (relative to the Dirac point of the kagome band structure in the non-interacting case). In particular, for $E_F = 0.2$ eV, the system is in a metallic regime, away from half-filling and from the Mott insulating regime, similar to the experimental measurements performed at the wire region of the hBN/Cu(111) moiré pattern (see Supplementary Fig. 2b). Note that in this metallic regime, our DMFT

calculations of $\Sigma(i\omega_n)$ converge by only considering low Matsubara frequencies $\omega_n$; this facilitates calculations at low temperatures [7]. Supplementary Fig. 3 shows the imaginary part of the DMFT-calculated self-energy $\Sigma(i\omega_n)$ as a function of $\omega_n$. We find that $\text{Im}[\Sigma(i\omega_{n=0})] \approx -0.00429$ eV, giving a mean-free path for a typical quasiparticle of $\ell \gtrsim 10$ nm.

# Supplementary Note 4 DMFT calculations: spectral functions $A(E)$ for different electron fillings

In Fig. 3 of the main text, the $DCA_3Cu_2$ MOF is metallic at the wire regions of the hBN/Cu(111) moiré pattern, but with an energy gap at the Fermi level elsewhere. In Fig. 4 of the main text, however, we show that for large tip-sample distances the MOF also appears to be in the Mott insulating regime at the wire region. This difference is related to the different tip-sample distances used in these two figures.

In Fig. 3 of the main text, we used a tip-sample distance, $\Delta z + z_0$, where $z_0$ is defined by the STM setpoint of $V_b = 10$ mV, $I_t = 10$ pA, and $\Delta z = 190$ pm. The largest tip-sample distance used in Fig. 4 of the main text, however, consists of $\Delta z = 330$ pm, with $z_0$ defined by the same setpoint. Importantly, Fig. 4e of the main text shows that, at the wire region, the MOF is in a gap-less metallic phase for a $\Delta z$ range between 175 and 275 pm; the tip-sample distance used in Fig. 3 of the main text is within this range. Based on Fig. 4e of the main text (which captures the energy shifts of $dI/dV$ features with changing $\Delta z$), we estimate that the chemical potential $E_F$ could shift by $\sim$45 meV from $\Delta z = 330$ pm to $\Delta z = 190$ pm.

Figure 3d of the main text shows the DMFT-calculated spectral functions, $A(E)$, with a chemical potential $E_F$ modulated by 0.2 eV, to match the expected work function modulation resulting from the hBN/Cu(111) moiré pattern. The agreement between the calculated spectral functions and the experimental $dI/dV$ spectra in Fig. 3b provides compelling evidence that the theory captures the fundamental physics observed experimentally. To capture the potential effect of the tip-sample distance difference, we performed the same DMFT calculations of $A(E)$ as in Fig. 3d of the main text, but with $E_F$ shifted upwards by 45 meV (Supplementary Fig. 4). Compared to Fig. 3d of the main text, these spectral functions are all gapped at the Fermi level and in the Mott insulating regime, even for the smallest $E_F$. This is consistent with the experimental spectra acquired at the wire region with large tip-sample distances (Fig. 4d of the main text), and further corroborates our theoretical model of the Mott insulating $DCA_3Cu_2$ MOF.

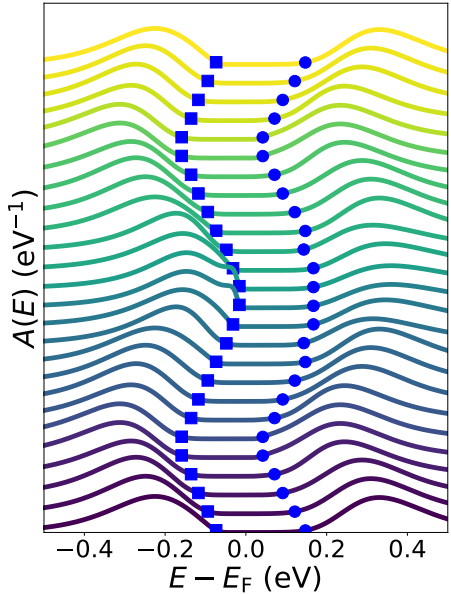

**Supplementary Fig. 4**: Spectral functions $A(E)$ calculated by DMFT, with 200 meV modulation of the chemical potential, $E_F$, to match the experimental hBN/Cu(111) moiré work function modulation. This calculation is the same as in Fig. 3 of the main text, but with $E_F$ shifted upwards by 45 meV for all spectra to capture the influence of different tip-sample distances and of tip-induced gating. In this case, all spectra remain gapped at the Fermi level.

# Supplementary Note 5 DMFT calculations: k-resolved spectral function

To further investigate the correlated-electron Mott insulating phase of the $DCA_3Cu_2$ MOF, we calculated via DMFT the $k$-dependent spectral function, $A(E, \mathbf{k}) = \mathrm{Im}\left[-1/\left[\pi(E + E_F - \varepsilon(\mathbf{k}) - \Sigma(E))\right]\right]$, for $U = 0.65$ eV and $E_F = 0.4$ eV (corresponding to ~half-filling of the three near-Fermi MOF kagome bands), where $\varepsilon(\mathbf{k})$ is the non-interacting kagome band structure and $\Sigma(E)$ is the self-energy analytically continued to real energy (Supplementary Fig. 5).

We observe a significant gap at the Fermi level ($E = 0$), with an occupied lower Hubbard band (LHB) with weakly dispersive features (in particular close to the $\Gamma$ point), and an empty upper Hubbard band (UHB) with significantly less dispersion. The *diffuse* nature of these bands (as opposed to sharp and well-defined, as in conventional band theory) is indicative of band incoherence; this is expected for a Mott insulating phase. The integration of $A(E, \mathbf{k})$ over $\mathbf{k}$ results in the spectral function $A(E)$ shown and discussed throughout the manuscript (e.g., orange curve in Fig. 1e). It is important to

note that DMFT can depict correlated-electron phenomena (including Mott metal-insulator transitions) more reliably and accurately than DFT [8–10].

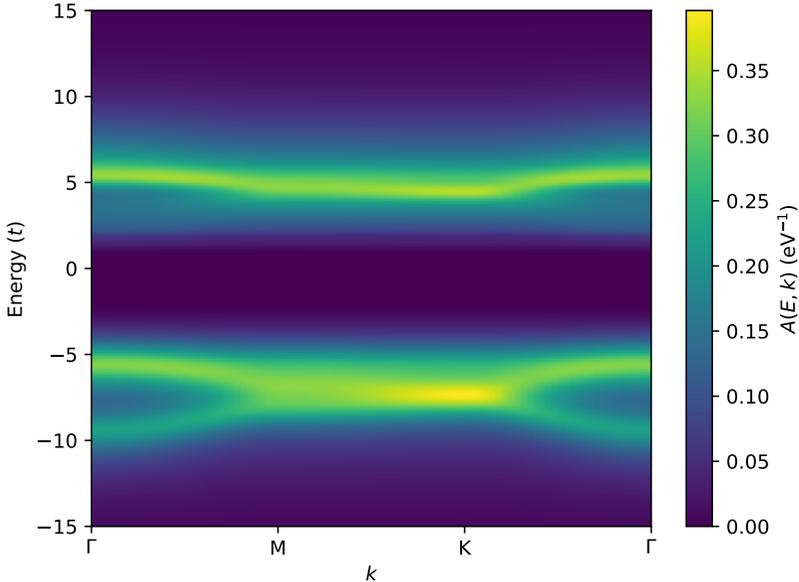

**Supplementary Fig. 5**: $k$-resolved spectral function for freestanding DCA$_3$Cu$_2$ MOF, calculated by DMFT ($U = 0.65$ eV, $t = 0.05$ eV, $E_\mathrm{F} = 0.4$ eV, corresponding to half-filling of the near-Fermi kagome bands).

# Supplementary Note 6  Growth and commensurability

In initial preparation attempts of DCA$_3$Cu$_2$ on hBN/Cu(111), we performed subsequent depositions of DCA molecules and then Cu atoms, as reported in the Methods section of the main text, but with the substrate held at room temperature (RT) throughout. Characterisation of these surfaces revealed supramolecular arrays of purely DCA molecules (that is, without DCA-Cu coordination) as the dominant phase, consistent with previous findings [11], with only small regions of MOF (Supplementary Fig. 6a-b). We found that, by first depositing DCA with the substrate held at RT, and subsequently depositing Cu with the substrate cooled to ~77 K, a greater coverage of MOF could be achieved. This low temperature method, however, also stabilised other DCA-Cu coordination motifs, leading to other amorphous structures, e.g., see Supplementary Fig. 6c, d. Further subsequent annealing of the sample

to ∼200 K led to monocrystalline DCA$_3$Cu$_2$ MOF domains as the most abundant phase, with only small regions of the supramolecular DCA-only phase (Supplementary Fig. 6e). There are some defects (cracks, vacancies, Cu clusters) within the MOF domain (see Supplementary Note 11 for discussion on effects of such defects on the MOF electronic properties).

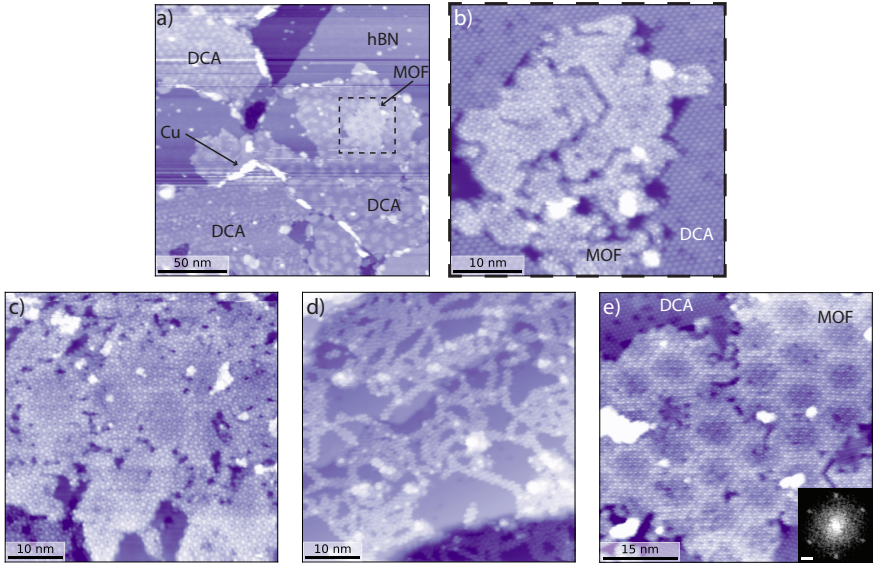

**Supplementary Fig. 6**: **Growth of DCA$_3$Cu$_2$ MOF on hBN/Cu(111).** **a,** Constant-current STM image showing mostly supramolecular DCA-only domains [11], with one small region of DCA$_3$Cu$_2$ MOF (dashed box), after sequential deposition of DCA then Cu with the substrate held at RT ($V_b = 1$ V, $I_t = 10$ pA). **b,** Smaller scale constant-current STM image showing the region of MOF highlighted by dashed box in (a) ($V_b = -1$ V, $I_t = 10$ pA). **c,** Constant-current STM image showing larger regions of DCA$_3$Cu$_2$ MOF after sequential deposition of DCA with the sample at RT and then Cu with the sample at ∼77 K ($V_b = -1$ V, $I_t = 10$ pA). **d,** Constant-current STM image showing amorphous structures DCA-Cu coordination structures stabilized by the same preparation method as in (c). **e,** Constant-current STM image showing a monocrystalline DCA$_3$Cu$_2$ MOF domain after annealing the sample shown in (c, d) to ∼200 K ($V_b = -1$ V, $I_t = 10$ pA). Inset: Fourier transform of STM image (scalebar 0.25 nm$^{-1}$).

Supplementary Fig. 7 shows a large-scale STM image (size: 300x300 nm$^2$) of the DCA$_3$Cu$_2$ MOF on hBN/Cu(111). We do not observe any DCA-only regions across this large-scale image. There is some remaining excess Cu on

the surface (as labelled). This indicates that, although not perfect, the stoichiometry for this sample preparation was close to the required 3:2 DCA:Cu, over a large sample region. These results are similar to prior work involving growth of this MOF on graphene [12].

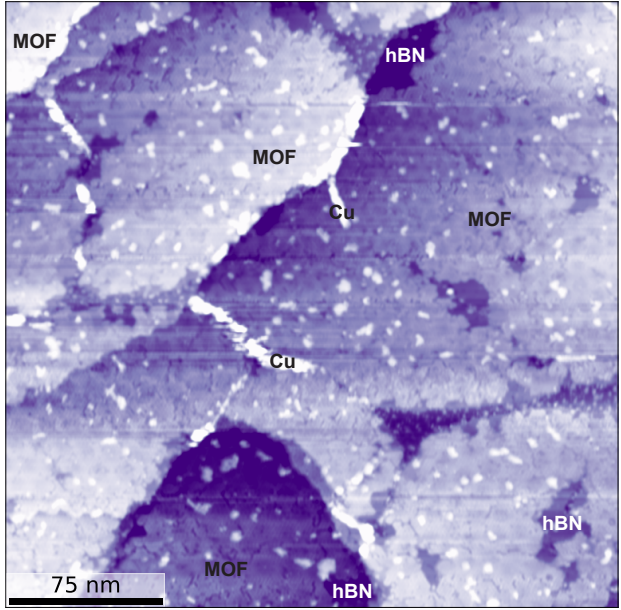

**Supplementary Fig. 7**: **Large-scale STM image of DCA$_3$Cu$_2$ MOF on hBN/Cu(111).** Image size: 300x300 nm$^2$ ($V_b = -1$ V, $I_t = 10$ pA). We observe no DCA-only regions, with only minor amounts of excess Cu. The sample preparation was close to the desired 3:2 DCA:Cu stoichiometry.

The DCA$_3$Cu$_2$ MOF was found to have a hexagonal lattice constant of 2.01 $\pm$ 0.06 nm. This suggests it is commensurate with the underlying hBN which is known to have a hexagonal lattice constant of 0.25 nm. The MOF was found to be incommensurate, however, with the hBN/Cu(111) moiré pattern. This is shown in the fast Fourier transform (FFT) in Supplementary Fig. 8b, c of the constant-current STM image in Supplementary Fig. 8a. The hBN/Cu(111) moiré pattern in this image has a lattice constant $\lambda = 10.7 \pm 0.9$ nm, at an angle of $32.0 \pm 0.5°$ with respect to the MOF lattice. The MOF appears to seamlessly grow across moiré domains of different sizes (due to different hBN domain rotations); see Supplementary Fig. 8d, and Figs. 3 and 4 of main text. This is further evidence that the MOF is incommensurate with the hBN/Cu(111) moiré pattern. Establishing whether the DCA$_3$Cu$_2$ MOF preferentially grows on hBN domains with a particular moiré periodicity is beyond the scope of this work and requires further investigation.

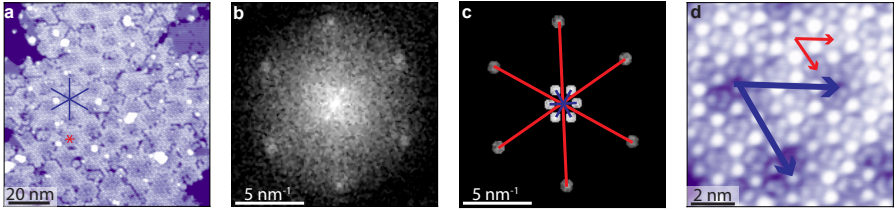

**Supplementary Fig. 8: Lattice vectors of DCA₃Cu₂ MOF on hBN/Cu(111), and of hBN/Cu(111) moiré pattern. a,** 100x100 nm² constant-current STM image of MOF on hBN/Cu(111) with moiré pattern clearly visible ($V_b = -1$ V, $I_t = 10$ pA). Hexagonal MOF (moiré) lattice vectors shown in red (blue, respectively). **b,** FFT of image in (a). **c,** Filtered version of FFT in (b), emphasising periodic features of MOF and hBN/Cu(111) moiré lattices. The MOF lattice vectors have a length of 2.01 ± 0.06 nm, at an angle of 32° with respect to the moiré lattice vectors which have a length of 10.7± 0.9 nm. **d,** Constant-current STM image of MOF on hBN/Cu(111) moiré domain with a smaller period of ∼5 nm ($V_b = -1$ V, $I_t = 10$ pA).

# Supplementary Note 7 Experimental d$I$/d$V$ vs. DMFT spectral functions

Supplementary Fig. 9 shows a comparison between STS measurements performed on the DCA₃Cu₂ MOF within a pore region of the hBN/Cu(111) moiré pattern (as shown in Fig. 2 of the main text), and a DMFT spectral function (for a corresponding value of chemical potential $E_F$). Note that the STS measurements were performed at 4 K, whereas the DMFT calculations were performed at a temperature of 29 K (for computational convenience) and hence appear broader.

The two spectra in Supplementary Fig. 9a are asymmetric in energy: in the DCA lobe site spectrum, the UHB signature is stronger than the LHB signature; in the Cu site spectrum, the LHB signature is stronger. This may be indicative of differences in the spatial distribution of electronic states associated with these two bands. The DMFT spectral function (Supplementary Fig. 9c) does not show such asymmetry, as DMFT is unable to capture any spatially dependent orbital texture of the corresponding electronic states. The orbital texture of electronic states of Mott insulating phases is known to be poorly captured by theoretical methods in any case [13]. Note that the sum of the two d$I$/d$V$ spectra in Supplementary Fig. 9a (Supplementary Fig. 9b) appears more symmetric, in agreement with the DMFT calculation.

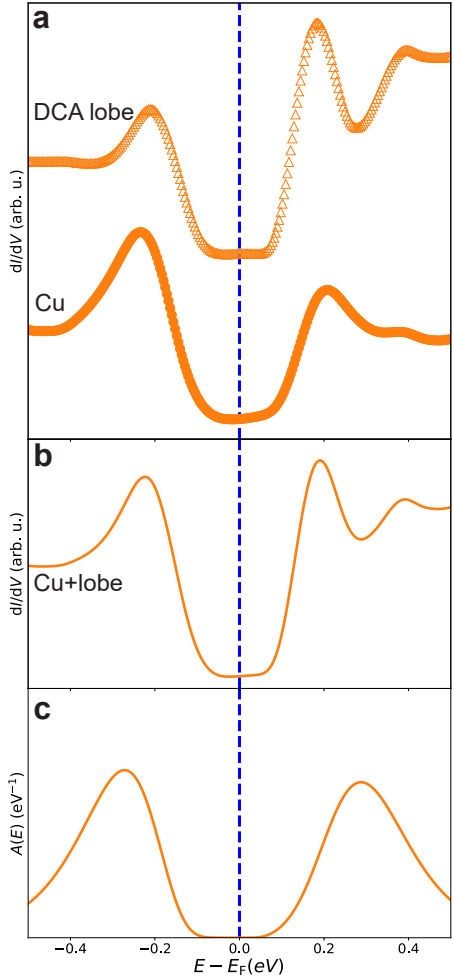

**Supplementary Fig. 9**: **Comparison between experimental** d$I$/d$V$
**spectra and DMFT spectral function for DCA$_3$Cu$_2$ MOF at a**
**hBN/Cu(111) moiré pore region. a,** Experimental d$I$/d$V$ spectra, as
shown in Fig. 2a of main text. Upper (lower) curve acquired at DCA lobe (Cu,
respectively) site (tip-sample distance defined by setpoint $V_b$ = -500 mV, $I_t$
= 500 pA). **b,** Sum of the two spectra in (a). **c,** Spectral function, $A(E)$, pre-
dicted by DMFT, as shown in Fig. 1e of main text [$U$ = 0.65 eV; $E_F$ chosen to
match electron filling predicted by DFT for MOF on hBN/Cu(111)]. Asym-
metry of spectra in (a) with respect to energy may be indicative of differences
in spatial distribution of electronic states associated with LHB and UHB.

Similarly, Supplementary Fig. 10 shows a comparison between STS mea-
surements performed on the DCA$_3$Cu$_2$ MOF at a wire region of the

hBN/Cu(111) moiré pattern (as shown in Fig. 4 of the main text; for a large tip-sample distance), and a DMFT spectral function (for a corresponding value of $E_F$). Note the charging feature at positive bias voltage in the experimental data (due to the proximity of the LHBM to the Fermi level in the case of the wire region; see main text), which is not captured by the DMFT calculation.

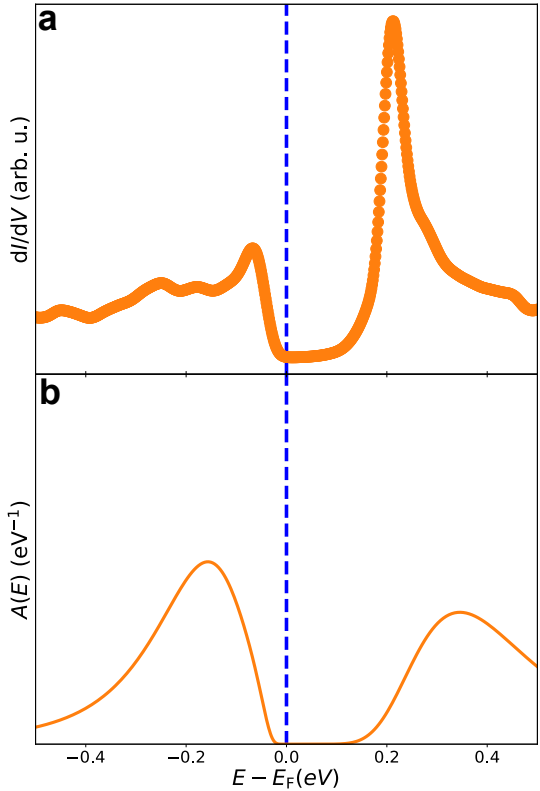

**Supplementary Fig. 10**: **Comparison between experimental $dI/dV$ spectrum and DMFT spectral function for $DCA_3Cu_2$ MOF at a moiré wire region. a,** Experimental $dI/dV$ spectrum acquired at Cu site of the MOF within wire region of hBN/Cu(111) moiré pattern (large tip-sample distance; see Fig. 4 of main text). Tip shifted 310 pm away from surface with respect to setpoint defined by $V_b = 10$ mV, $I_t = 10$ pA. **b,** Spectral function, $A(E)$, calculated by DMFT ($U = 0.65$ eV; $E_F$ chosen to match LHBM position of experimental spectrum in panel a).

# Supplementary Note 8  Bias-dependent STM imaging

We show a representative sample of bias-dependent constant-current STM images in the main text Fig. 2. Supplementary Fig. 11 shows a full set of these images acquired from $V_b = -1$ V to 0.6 V. For $V_b \lesssim$ -200 mV and $V_b \gtrsim$ 200 mV, the MOF appearance remains relatively unchanged, demonstrating the characteristic spatial distribution of the electronic states contributing to the $DCA_3Cu_2$ kagome bands, that is, with strong intensity at the Cu sites and the ends of the DCA anthracene moieties. At bias voltages closer to the Fermi level, however, the MOF appearance changes and shows fairly uniform intensity across the DCA molecules, with reduced intensity at the Cu sites, indicative of topographic rather than electronic features. This provides strong evidence that the $dI/dV$ peaks at $\sim \pm 200$ mV in Fig. 2 of the main text are related to intrinsic $DCA_3Cu_2$ electronic states. Variation in imaging across Supplementary Fig. 11 is attributed to the influence of the hBN/Cu(111) moiré pattern, as shown in Fig. 3 of the main text.

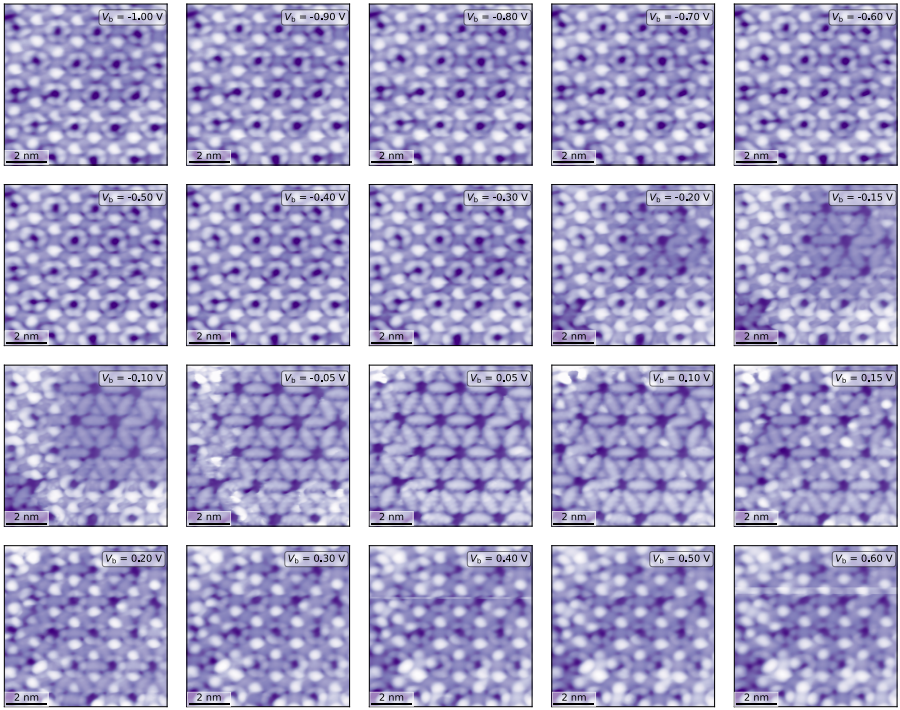

**Supplementary Fig. 11**: Constant-current STM images of $DCA_3Cu_2$ on hBN/Cu(111), at bias voltages specified in each image ($I_t = 10$ pA).

# Supplementary Note 9 d$I$/d$V$ maps

To depict the spatial distribution of $DCA_3Cu_2$ MOF electronic states at different energies, we show, in the main text Fig. 2, constant-current STM images at different bias voltages $V_b$, rather than d$I$/d$V$ maps. The spatial variation in electronic properties caused by the hBN/Cu(111) moiré pattern (main text Fig. 3) is reflected in both topographic images and d$I$/d$V$; see Supplementary Figs. 11 and 12. Constant-current STM images, which represent the integral of the local density of states (LDOS; or spectral function in general) between $eV_b$ and the Fermi level, are however less affected than the d$I$/d$V$ maps ($\propto$ LDOS) by these moiré-related variations. These constant-current STM images provide a good depiction of the intrinsic electronic character of the MOF in the electronically decoupled environment that the monolayer hBN provides.

Nonetheless, d$I$/d$V$ maps are consistent with the observed spatially dependent shifts in the energy onset of bands, shown in Fig. 3 of the main text. Illustrative examples of d$I$/d$V$ maps, acquired as slices of the numerical derivative of pixel-by-pixel $I(V)$ curves, are shown in Supplementary Fig. 12. These clearly show strong d$I$/d$V$ intensity associated with the maxima of lower Hubbard bands (LHB) at the Cu and anthracene end sites of the $DCA_3Cu_2$ MOF (as shown in Fig. 2 of the main text). This intensity corresponding to the LHB appears throughout the entire hBN/Cu(111) moiré pattern at $V_b = -0.30$ V (Supplementary Fig. 12b), including at the centre of the pore region. At $V_b$ closer to the Fermi level, however, the pore becomes progressively darker as spectra are acquired in the bandgap of the LDOS (or spectral function) in these regions.

At positive bias voltages, a similar trend is observed for the upper Hubbard band (UHB), with corresponding d$I$/d$V$ intensity at the DCA lobes and Cu sites of the MOF throughout the entire hBN/Cu(111) moiré pattern at $V_b = 0.30$ V (Supplementary Fig. 12m), with the pore regions becoming progressively darker as $V_b$ is reduced towards the Fermi level due to the energy gap (Supplementary Fig. 12h-l). The trend is less obvious for positive biases than for negative, however, as the d$I$/d$V$ signal at wire regions show charging features in this energy range (as shown in Fig. 4 of main text).

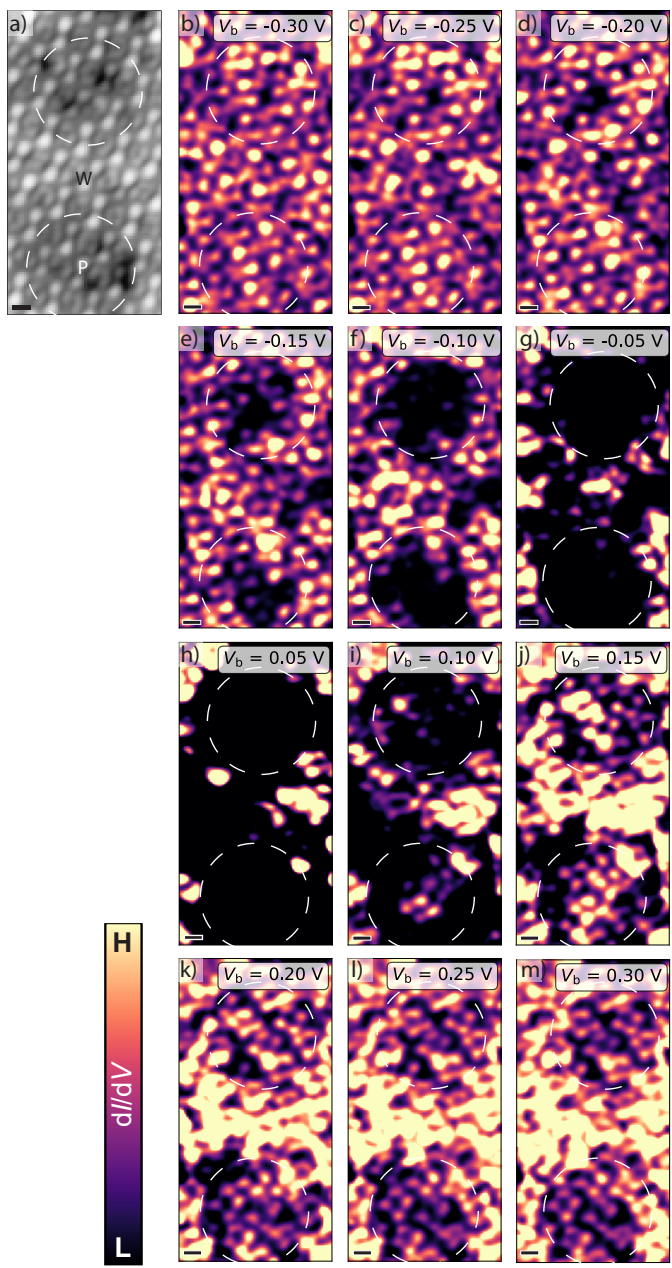

**Supplementary Fig. 12**: **d$I$/d$V$ maps of DCA$_3$Cu$_2$/hBN/Cu(111). a,** Constant-current STM image of DCA$_3$Cu$_2$/hBN/Cu(111) containing two pore (P) regions (dashed white circles) separated by a wire (W) region ($V_b = -1$ V, $I_t = 10$ pA). **b-m,** d$I$/d$V$ maps at the bias voltages specified in each panel, acquired in the same region as (a). At each pixel, the tip height was stabilised 225 pm further away from the surface with respect to a setpoint of $V_b = 10$ mV, $I_t = 10$ pA, before $I(V)$ acquisition. Scale bars: 1 nm.

# Supplementary Note 10 STS measurements at MOF high-symmetry points

In the main text, we focus on DCA lobe and Cu sites of the MOF to highlight its electronic properties, since the electronic states associated with the near-Fermi kagome bands have predominantly DCA LUMO character (with some Cu character), and such states are mostly localised at these sites [12, 14]. STS measurements at other high-symmetry locations within the MOF (e.g., DCA molecule centre; MOF pore site – to not confuse with hBN/Cu(111) moiré pore region) are displayed in Supplementary Fig. 13b. These MOF pore (yellow) and DCA centre (blue) spectra appear qualitatively consistent with the DCA lobe (green) and Cu (purple) site spectra, with a gap at the Fermi level between LHB and UHB. Note that the d$I$/d$V$ signal associated with LHB and UHB is weaker at MOF pore and DCA centre sites than at DCA lobe and Cu sites. The d$I$/d$V$ signal at MOF pore and DCA centre sites is given by signatures of the MOF near-Fermi electronic structure which are strongest at DCA lobe and Cu sites.

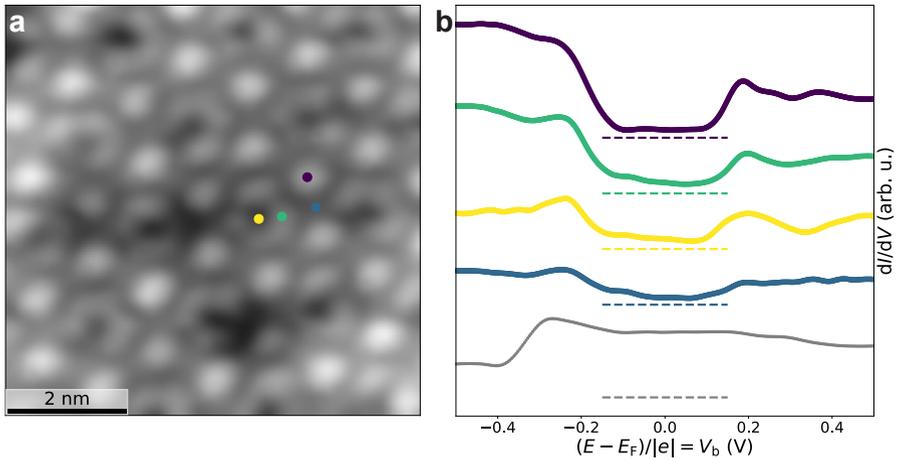

**Supplementary Fig. 13**: **STS measurements at MOF high-symmetry locations within a pore region of the hBN/Cu(111) moiré pattern. a,** STM image of DCA$_3$Cu$_2$ MOF on hBN/Cu(111) ($V_b = -1$ V, $I_t = 10$ pA). **b,** STS measurements performed at positions corresponding to coloured markers in (a). Grey curve: reference spectrum acquired upon bare hBN/Cu(111). Curves offset for clarity. Dashed lines indicate position of d$I$/d$V = 0$. Tip height stabilised 150 pm further away from the surface with respect to set-point of $V_b = 10$ mV, $I_t = 10$ pA.

# Supplementary Note 11 Electronic influence of defects

The $DCA_3Cu_2$ MOF domains on hBN/Cu(111) show good monocrystallinity, arguably given by the good match between MOF and hBN lattices. Nonetheless, a number of defects such as domain edges and boundaries, vacancies, and cracks can be observed (see main text Fig. 1a). To investigate the influence of these defects on the electronic properties of the MOF, we conducted STS measurements at the locations of these defects.

STS measurements performed at the edge of a $DCA_3Cu_2$ MOF domain (within a pore region of the hBN/Cu(111) moiré) are shown in Supplementary Fig. 14. At the very edge of the domain, DCA molecules are coordinated to only one Cu atom (rather than two as is the case within the bulk). Spectra acquired at these positions exhibit a LHB feature, similar to spectra within the MOF domain reported elsewhere in this manuscript. However, these spectra feature heavily reduced UHB weight and an additional peak at ∼0.5-0.7 V, not seen for spectra within the MOF domain. This additional peak is reminiscent of the DCA LUMO for purely organic films of DCA molecules on hBN/Cu(111) [11]. The peak observed here appears at a lower energy, however. We therefore claim that the DCA molecules at the edges of the MOF domain, coordinated to just one Cu atom, exhibit electronic properties which are a mixture of typical $DCA_3Cu_2$ properties and those of DCA-only domains. Importantly, just one unit cell away from the domain edges, the $DCA_3Cu_2$ properties are identical to those within a MOF domain region without defects, with clear LHB and UHB signatures, indicating that these domain edges do not significantly affect the electronic characteristics of the MOF domain bulk discussed throughout the manuscript.

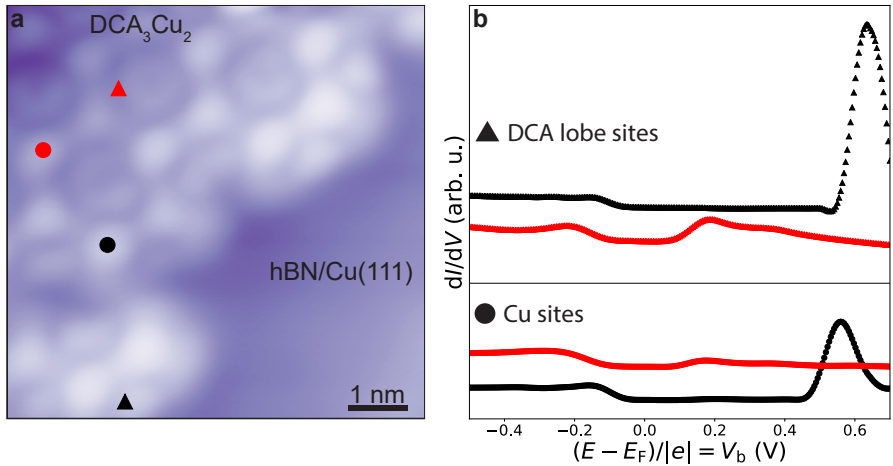

**Supplementary Fig. 14**: **STS measurements at the edge of a DCA₃Cu₂ MOF domain. a,** STM image of the edge of a MOF domain ($V_b = -1$ V, $I_t = 10$ pA). **b,** STS measurements at Cu sites (circle markers) and DCA lobe sites (triangle markers) both at the edge of the MOF domain (black) and within the MOF domain (red). Curves offset for clarity. Setpoints: $V_b = -500$ mV, $I_t = 100$ pA.

We observed similar electronic properties at boundaries between $DCA_3Cu_2$ MOF domains. These domain boundaries can be observed in a number of STM images presented in this work and appear as dark "cracks" within the MOF domains (Supplementary Fig. 15a). Similar to the edges of MOF domains, these domain boundaries feature DCA molecules coordinated to just one Cu atom. STS measurements performed at one of these domain boundaries are shown in Supplementary Fig. 15b. As was the case for the edge of the MOF domain, the spectrum acquired at the domain boundary exhibits reduced UHB weight and an additional peak at ~0.55 V. This again suggests that the electronic properties of the singly Cu-coordinated DCA molecules here are a mixture of Mott insulating $DCA_3Cu_2$ and the LUMO of the DCA-only phase.[11] Once again, just one unit cell away from these boundaries, the electronic properties are identical to those within a MOF domain region without defects, indicating that domain boundary defects do not significantly affect the intrinsic electronic properties of a MOF domain bulk.

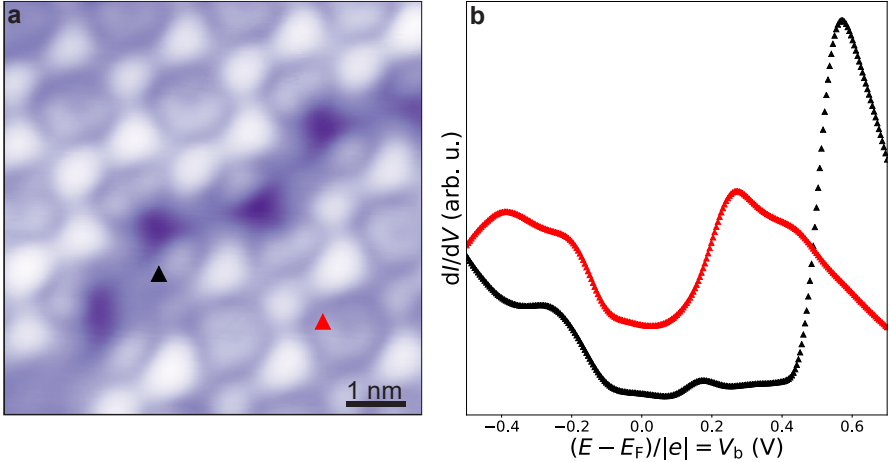

**Supplementary Fig. 15**: **STS measurements at DCA$_3$Cu$_2$ MOF domain boundary. a,** STM image showing a MOF domain boundary ($V_b = -1$ V, $I_t = 10$ pA). **b,** STS measurements at DCA lobe sites, both at the MOF domain boundary (black) and within the MOF domain away from the boundary (red). Curves offset for clarity. Setpoints: $V_b = -500$ mV, $I_t = 100$ pA.

Finally, we also considered the influence of Cu vacancy defects which were present in some MOF domains. Two spectra at Cu sites of the DCA$_3$Cu$_2$ MOF in proximity to a Cu vacancy (green and orange), and one spectrum at the Cu vacancy site (blue) are shown in Supplementary Fig. 16. The three spectra are qualitatively similar, featuring a pronounced energy gap across the Fermi level indicating the Mott insulating phase. The blue spectrum, however, has noticeably weaker signal at the LHB and UHB. Overall, the vacancy defect does not appear to significantly perturb the MOF's electronic properties.

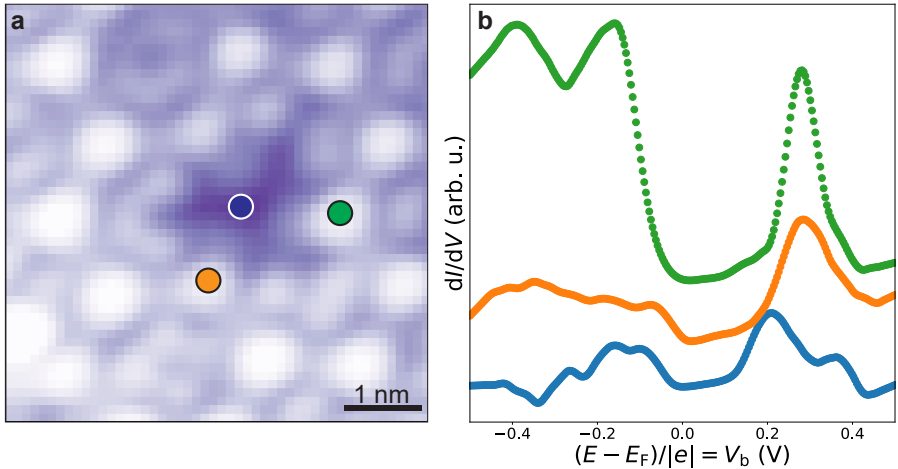

**Supplementary Fig. 16**: **STS measurements at Cu vacancy defect within a DCA$_3$Cu$_2$ MOF domain. a,** STM image showing a Cu vacancy defect ($V_b = -1$ V, $I_t = 10$ pA). **b,** STS measurements at Cu sites in proximity of Cu vacancy (green and orange), and at Cu vacancy (blue). Curves offset for clarity. Setpoints: 135pm further away from the surface with respect to a setpoint of $V_b = 10$ mV, $I_t = 10$ pA.

# Supplementary Note 12 Broad energy-range STS measurements

In the main text, we focus on STS measurements in an energy window from $-0.5$ to $0.5$ eV. This energy window was selected based on DFT predictions according to which the near-Fermi kagome MOF bands are well isolated and separated in energy ($>1$ eV) from other bands. To verify this assumption, we conducted STS measurements over a broader energy window, from -1 V to 1 V, at a DCA lobe site of the DCA$_3$Cu$_2$ MOF within a pore region of the hBN/Cu(111) moiré pattern (Supplementary Fig. 17). The LHB and UHB features, as well as the $\sim$200 meV energy gap, are evident in the MOF spectrum. No other significant features can be observed in this energy window. This is consistent with the theoretical predictions and justifies our focus on the near-Fermi energy window within the main text.

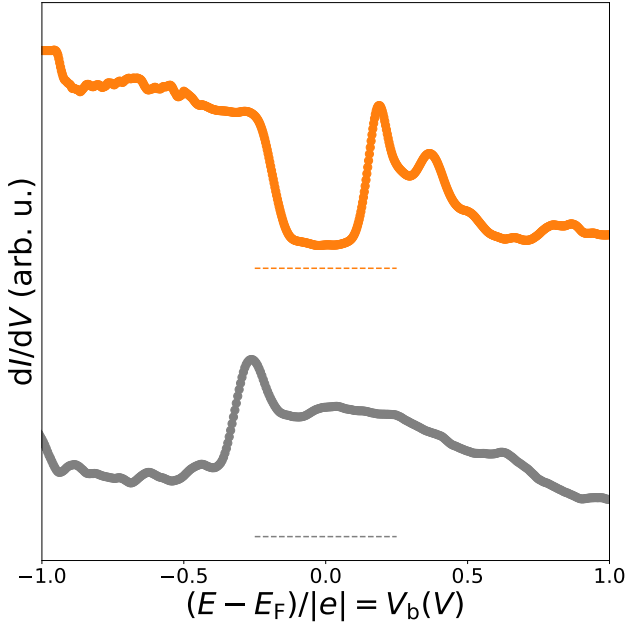

**Supplementary Fig. 17**: **STS measurements over broad energy window.** Orange curve: DCA lobe site of MOF within pore region of hBN/Cu(111) moiré pattern. Grey curve: bare hBN/Cu(111) reference spectrum. Spectra normalised and offset for clarity. Orange curve acquired in two parts: setpoint of $V_b = -1$ V, $I_t = 100$ pA from -1 V to -500 mV; setpoint of $V_b = -500$ mV, $I_t = 100$ pA for the remaining bias range. Grey curve setpoint: $V_b = -2$ V, $I_t = 100$ pA.

# Supplementary Note 13 Determination of band edges

In Figs. 3 and 4 of the main text, MOF band edges are tracked as a function of position on the hBN/Cu(111) moiré pattern and of tip-sample distance, respectively. Determining precisely the bias voltage $V_b$ corresponding to these band edges can be challenging due to a non-zero $dI/dV$, even in the gap region, resulting from the underlying Cu(111) LDOS.

We therefore determined such band edges by finding a $V_b$ value, $V_b^*$, corresponding to a local minimum of $dI/dV$ in a small window about $V_b = 0$ (i.e., $d^2I/dV^2 = 0$). We then determined the lower (upper) Hubbard band edge as the value of $V_b$ closest to and below (above) $V_b^*$, for which $|d^2I/dV^2(V_b)| > \epsilon$, where $\epsilon$ represents a specified threshold. A consequence of this method was that lower Hubbard band edges were sometimes located at $V_b > 0$ V; we claim this is consistent with the observed reduction of the hole creation energy barrier at the wire regions of the hBN/Cu(111) moiré pattern (suggesting that the

Mott phase collapses in these regions, consistent with the spectral functions predicted by DMFT for different values of $E_F$ in Supplementary Fig. 2). An example of the application of this method is shown in Supplementary Fig. 18.

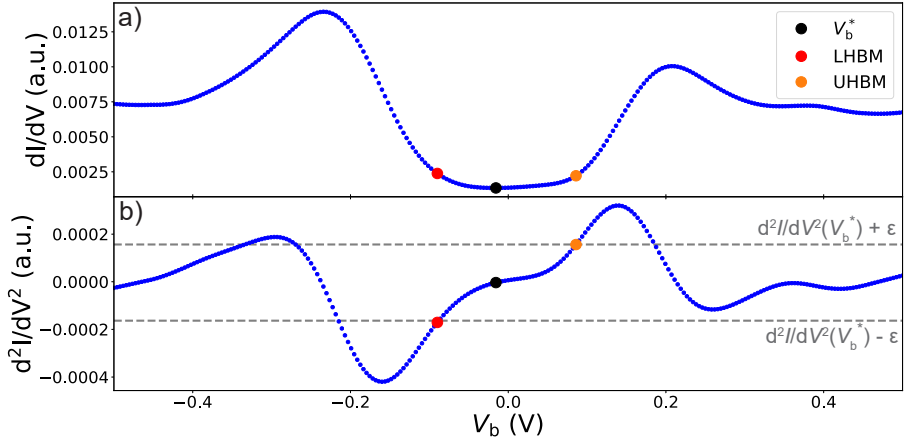

**Supplementary Fig. 18**: **Method for determining band edges in d$I$/d$V$ spectra. a,** d$I$/d$V$ spectrum obtained via the numerical derivative of an $I(V)$ curve at a Cu site of DCA$_3$Cu$_2$ MOF on hBN/Cu(111) (setpoint: $V_b = -500$ mV, $I_t = 500$ pA). **b,** d$^2I$/d$V^2$ spectrum for the same data as in (a). Band edges are determined by the values of $V_b$ near $V_b^*$, a local minimum of d$I$/d$V$ near the Fermi level, for which $|d^2I/dV^2(V_b)| > \epsilon$ (where $\epsilon$ is a specified threshold).

The determination of band edges for the DMFT spectral functions $A(E)$ was simpler, since $A(E) \to 0$ for energies $E$ within the bandgap. This energy gap was identified as a continuous energy region around the chemical potential $E_F$ associated with values of the (normalised) spectral function smaller than some threshold value. The threshold was chosen as $5 \times 10^{-3}$ $t^{-1}$ (0.1 eV$^{-1}$). This value set the band edges at energy positions agreeing with those determined by the above method for experimental spectra.

# Supplementary Note 14 Moiré-dependent STS measurements at DCA lobe sites of DCA$_3$Cu$_2$ MOF

In Supplementary Fig. 3 of the main text, we showed STS measurements at Cu sites of the DCA$_3$Cu$_2$ MOF, at different locations of the hBN/Cu(111) moiré pattern, from pore to wire to pore regions. Equivalent measurements performed at the same hBN/Cu(111) moiré pattern as in Fig. 3, with similar tunneling parameters (e.g., intermediate tip-sample distances), but at DCA lobe sites

are shown in Fig. 19. This figure shows the same trend as Fig. 3. The LHB and UHB shift in energy across the electronically corrugated hBN/Cu(111) moiré pattern. At the wire regions, the energy gap of $\sim$200 meV at the Fermi level is no longer observed, and the Fermi-level $\mathrm{d}I/\mathrm{d}V$ signal increases. These findings are consistent with our interpretation that the $DCA_3Cu_2$ MOF is a Mott insulator on the hBN/Cu(111) substrate, and that a combination of template- and tip-induced gating can depopulate the kagome bands to induce a local Mott metal-insulator transition.

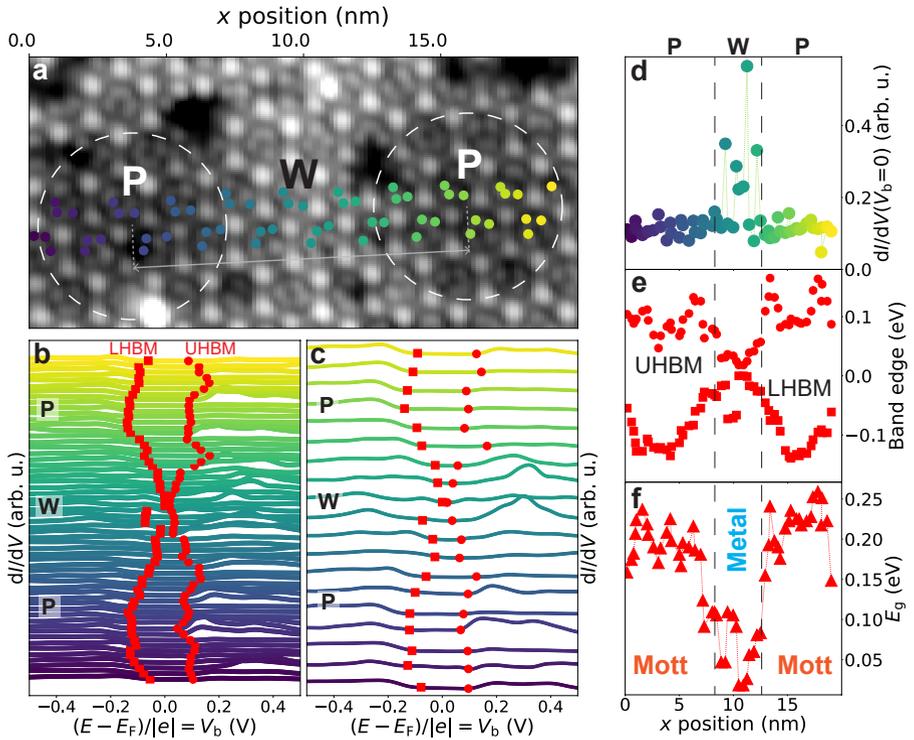

**Supplementary Fig. 19**: **STS measurements at DCA lobe sites of DCA$_3$Cu$_2$ MOF at different locations of the hBN/Cu(111) moiré pattern. a,** STM image of MOF ($V_b$ = -1 V, $I_t$ = 10 pA). White dashed circles (P): hBN/Cu(111) moiré pore regions, separated by wire region (W). Grey arrow indicates moiré period $\lambda \approx 12.5$ nm. **b,** d$I$/d$V$ spectra acquired at MOF DCA lobe sites, at positions indicated by coloured markers in (a). Tip 190 pm further from surface with respect to setpoint $V_b$ = 10 mV, $I_t$ = 10 pA). Energy gap $E_g \approx 200$ meV at Fermi level at P regions, with no Fermi-level gap at W region (LHBM: lower Hubbard band maximum; UHBM: upper Hubbard band minimum). **c,** Selection of spectra in (b), plotted for clarity. **d,** d$I$/d$V$ signal at Fermi level ($V_b$ = 0) as a function of position $x$ with respect to hBN/Cu(111) moiré pattern, for each spectrum in (b), showing increased signal within the W region. **e, f,** LHBM (squares), UHBM (circles) and energy gap $E_g$ (triangles), as a function of $x$ position, extracted from (b).

# Supplementary Note 15 STS measurements for hBN/Cu(111) moiré domains with different periods

Due to different rotational domains of hBN on Cu(111), moiré patterns with different periods are possible. As described in Supplementary Note 6, the $DCA_3Cu_2$ MOF grows seamlessly across moiré patterns with different periods.

To assess the influence of different hBN/Cu(111) moiré periods on the electronic properties of the $DCA_3Cu_2$ MOF, we performed STS measurements equivalent to those shown in Fig. 3 of the main text (for a hBN/Cu(111) moiré domain with a period $\lambda = 12.5$ nm) on two additional moiré patterns with periods of $\lambda \approx 10$ nm and $\lambda \approx 5$ nm (Supplementary Figs. 20, 21).

As in Fig. 3 of the main text, both Supplementary Figs. 20 and 21 show the LHB and UHB shifting in energy as a function of position on the electronically corrugated moiré pattern. The LHBM varies sinusoidally from pore to wire to pore region, with the period of the moiré pattern (Supplementary Figs. 20d, 21d). These spectra are gapped at the Fermi level at the pore regions (therefore insulating), with no gap at the Fermi level (Supplementary Figs. 20e, 21e) and a pronounced increase in Fermi-level $dI/dV$ signal at the wire regions (therefore metallic; Supplementary Figs. 20c, 21c).

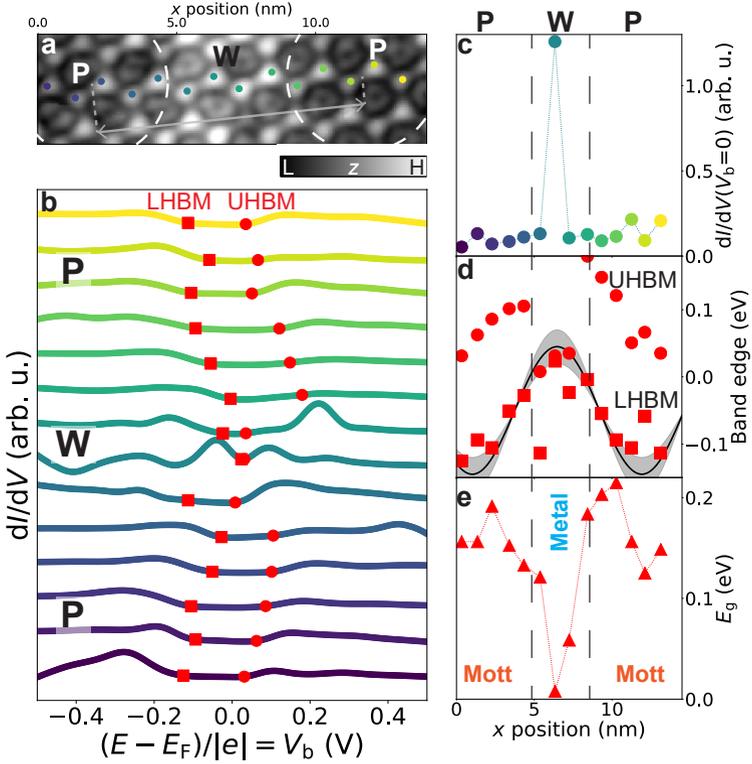

**Supplementary Fig. 20**: **STS measurements at Cu sites of MOF on hBN/Cu(111) domain with moiré period $\lambda \approx 10$ nm. a,** STM image of MOF ($V_\mathrm{b}$ = -1 V, $I_\mathrm{t}$ = 10 pA). White dashed circles (P): hBN/Cu(111) moiré pore regions, separated by wire region (W). Grey arrow indicates moiré period. **b,** $\mathrm{d}I/\mathrm{d}V$ spectra acquired at MOF Cu sites, at positions indicated by coloured markers in (a). Tip 190 pm further from surface with respect to setpoint $V_\mathrm{b}$ = 10 mV, $I_\mathrm{t}$ = 10 pA. Energy gap $E_\mathrm{g} \approx 200$ meV at Fermi level for P regions, vanishing at W region (LHBM: lower Hubbard band maximum; UHBM: upper Hubbard band minimum). **c,** $\mathrm{d}I/\mathrm{d}V$ signal at Fermi level ($V_\mathrm{b}$ = 0) as a function of position $x$ on hBN/Cu(111) moiré pattern, extracted from spectra in (b), showing increased signal within W region. **d, e,** LHBM (squares), UHBM (circles) and energy gap $E_\mathrm{g}$ (triangles) as a function of $x$ position, extracted from (b). Black line in (d): expected LHBM variation due to local moiré work function modulation for this moiré period [3]. Grey shaded area: uncertainty.

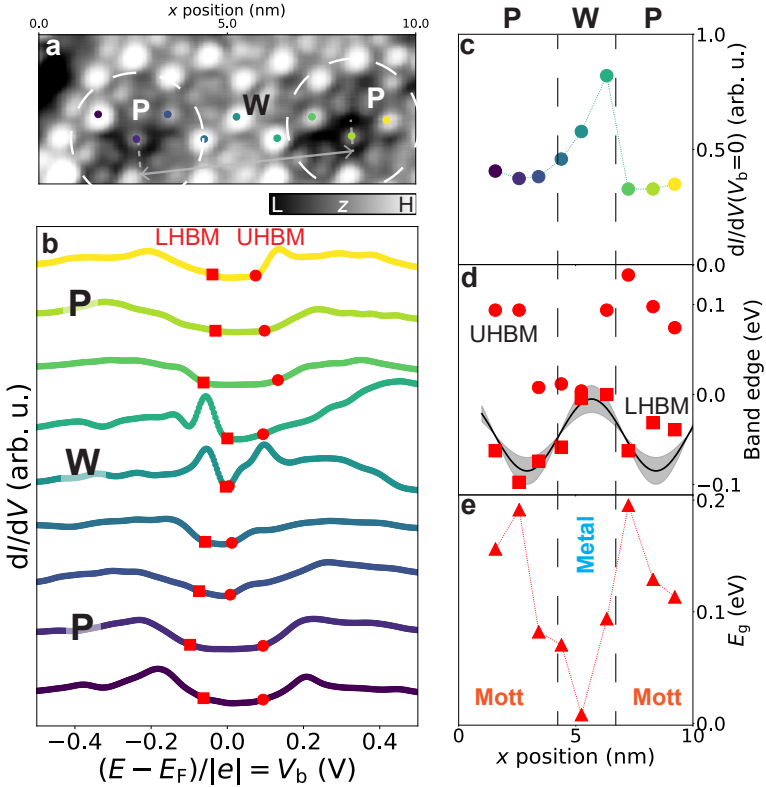

**Supplementary Fig. 21**: **STS measurements at Cu sites of MOF on hBN/Cu(111) domain with moiré period $\lambda \approx 5$ nm. a,** STM image of MOF ($V_b = -1$ V, $I_t = 10$ pA). White dashed circles (P): hBN/Cu(111) moiré pore regions, separated by wire region (W). Grey arrow indicates moiré period. **b,** $dI/dV$ spectra acquired at MOF Cu sites, at positions indicated by coloured markers in (a). Tip 190 pm further away from surface with respect to setpoint $V_b = 10$ mV, $I_t = 10$ pA. Energy gap $E_g \approx 200$ meV at Fermi level for P regions, vanishing at W region (LHBM: lower Hubbard band maximum; UHBM: upper Hubbard band minimum). **c,** $dI/dV$ signal at the Fermi level ($V_b = 0$) as a function of position $x$ on hBN/Cu(111) moiré pattern, extracted from spectra in (b), showing increased signal within W region. **d, e,** LHBM (squares), UHBM (circles) and energy gap $E_g$ (triangles) as a function of $x$ position, extracted from (b). Black line in (d): expected LHBM variation due to local moiré work function modulation for this moiré period [3]. Grey shaded area: uncertainty.

In Fig. 3 of the main text, the $dI/dV$ spectra with a gap at the Fermi level show a sinusoidal variation of the Hubbard band edges that follows the

variation of the local work function given by the hBN/Cu(111) moiré pattern. For a hBN/Cu(111) moiré pattern with a periodicity of ~12.5 nm (as in the main text), such local work function variation from pore to wire region is of ~200 meV [3]. For the two moiré patterns in Supplementary Figs. 20 and 21, the energy variation of the LHBM is smaller: ~190 meV for $\lambda \approx 10$ nm and ~100 meV for $\lambda \approx 5$ nm. These values are in good agreement with the local work function modulation for these moiré periods (black curves with grey uncertainty regions in Supplementary Figs. 20d and 21d) [3]. These findings are summarised in Supplementary Fig. 22. The black markers show the expected local work function modulation for different hBN/Cu(111) moiré periods [3]. The red markers show the experimentally observed variation in LHBM energy between pore and wire regions, for $\lambda \approx 10$ nm (Supplementary Fig. 20), $\lambda \approx 5$ nm (Supplementary Fig. 21) and $\lambda \approx 12.5$ nm (main text Fig. 3). Our experimental LHBM energy shifts lie within the uncertainty of the moiré local work function variation. This agreement validates our interpretation of the Hubbard band shifts resulting from the electronically corrugated hBN/Cu(111) moiré pattern. Notably, the electronic properties of the MOF are consistent for hBN/Cu(111) moiré domains with significantly different periods.

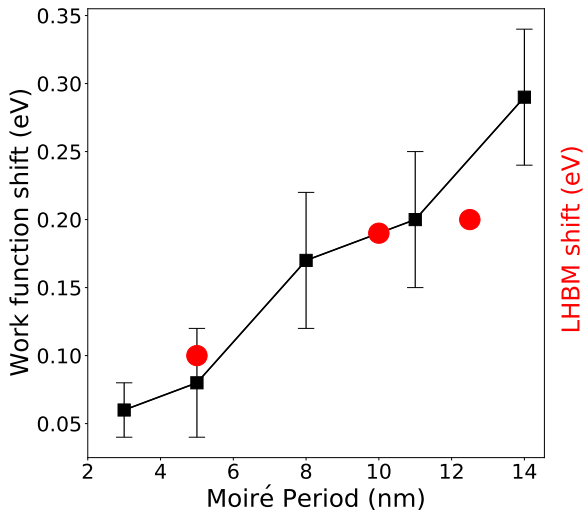

**Supplementary Fig. 22**: **Comparison between local work function modulation and LHBM variation for different moiré periods.** Black markers: local work function shift (and uncertainties) for different moiré periods, from prior literature [3]. Red markers: Experimentally observed variations in LHBM energy for the three different moiré periods considered in this work.

# Supplementary Note 16 Charging signatures

In the $dI/dV$ spectra in Fig. 4 of the main text, we observe narrow peaks whose energy position vary linearly as a function of tip-sample distance $\Delta z + z_0$. Based on a double-barrier tunnel junction (DBTJ) model, we attribute these features to charging of intrinsic electronic states of the MOF. Further evidence that these features are related to charging phenomena is provided by the presence of charging rings in $dI/dV$ maps. An example is shown in Supplementary Fig. 23. At a bias voltage $V_b \approx 0.05$ V (Supplementary Fig. 23b), we observe $dI/dV$ intensity at the centre of a MOF DCA molecule, within a wire region of the hBN/Cu(111) moiré pattern. As $V_b$ increases (Supplementary Fig. 23c-m), this $dI/dV$ intensity moves gradually towards the edges of the DCA molecule, with the shape of a ring, consistent with the expected charging ring behaviour [12, 15].

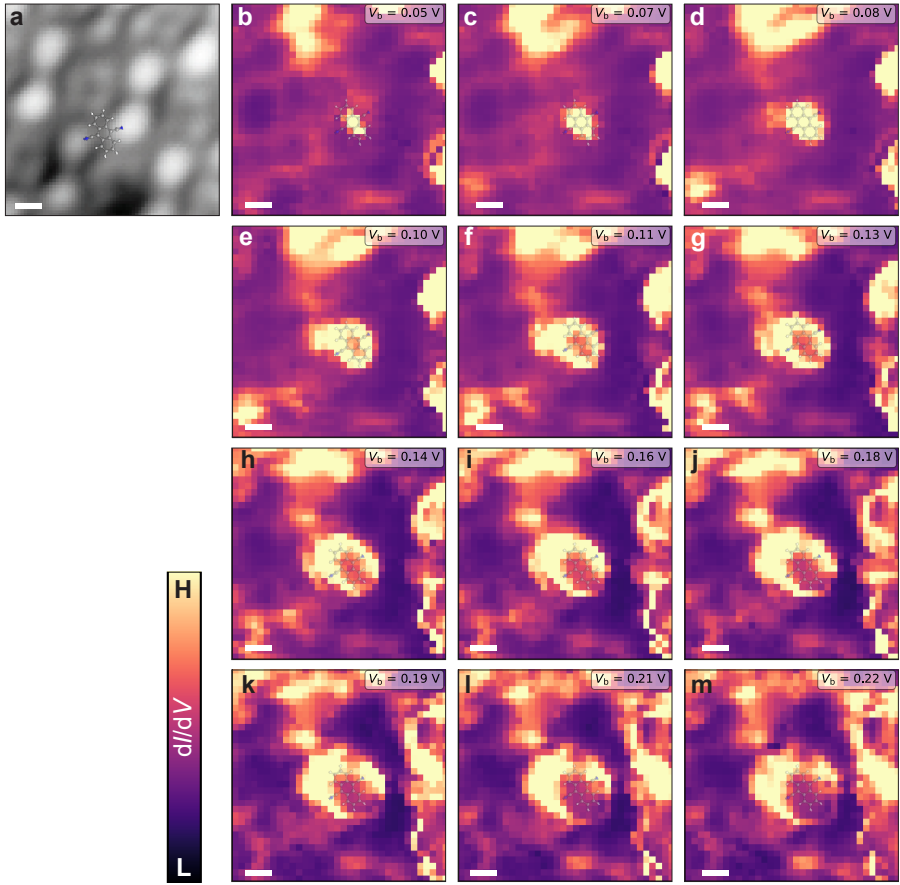

**Supplementary Fig. 23**: **d$I$/d$V$ maps of DCA$_3$Cu$_2$/hBN/Cu(111): charging ring. a,** STM image of DCA$_3$Cu$_2$ at a wire region of hBN/Cu(111) moiré pattern ($V_b = -1$ V, $I_t = 10$ pA). **b-m,** d$I$/d$V$ maps of region in (a), at indicated bias voltages $V_b$, obtained via numerical derivative of pixel-by-pixel $I(V)$ curves. At each pixel, the tip-sample distance was stabilised 300 pm further away from the surface relative to a setpoint of $V_b = 10$ mV, $I_t = 10$ pA, before $I(V)$ acquisition. Scale bars: 1 nm.

# Supplementary Note 17   Charging features in MOF d$I$/d$V$ at pore-wire boundary of moiré pattern

In Fig. 3b of the main text, some MOF d$I$/d$V$ spectra acquired at locations in between the hBN/Cu(111) moiré pore and wire regions featured peaks at biases of ~0.4 eV. These peaks are not captured by the corresponding DMFT calculations in Fig. 3d. We conducted further d$I$/d$V$ measurements as a function of

tip-sample distance $\Delta z + z_0$ at a MOF Cu site at such boundary locations of the moiré pattern (Supplementary Fig. 24). The $dI/dV$ spectra in Supplementary Fig. 24b exhibit a narrow peak at $V_b > 0.4$ V for intermediate $\Delta z$ values, progressively shifting to lower energies with decreasing $\Delta z$. This is very similar to the tip-induced charging behaviour due to the DBTJ effect demonstrated in Fig. 4. This provides an explanation for the DMFT spectral functions in Fig. 3d not showing such a peak, since DMFT does not capture tip-induced phenomena. Note that, for the range of tip-sample distances explored, we did not observe a transition to a metal-like phase at this specific acquisition site.

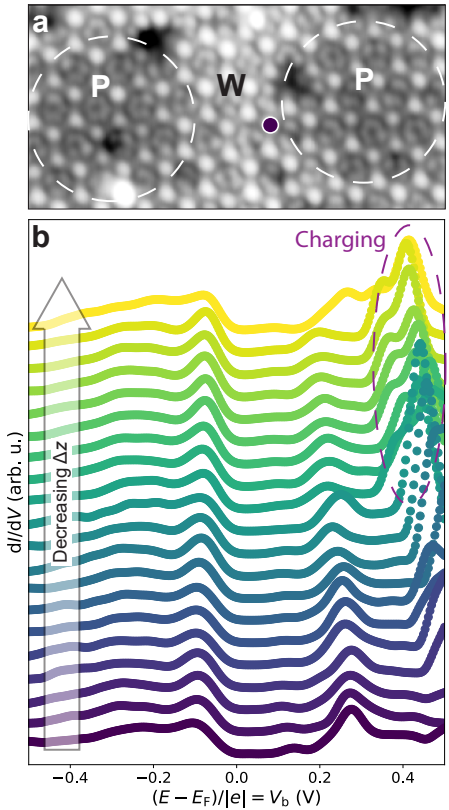

**Supplementary Fig. 24**: **Charging features at a pore-wire boundary of the moiré pattern. a,** STM image of MOF/hBN/Cu(111) showing two pore regions and one wire region of the hBN/Cu(111) moiré pattern ($V_b = -1$ V, $I_t = 10$ pA). **b,** $dI/dV$ spectra for different tip-sample distance changes $\Delta z$, acquired at MOF Cu site, as indicated in (a). Spectra normalised and offset for clarity. Setpoints range from 250 pm further from surface (bottom curve) to 105 pm further from surface with respect to a setpoint of $V_b = 10$ mV, $I_t = 10$ pA.

# Supplementary Note 18 Estimate of tip work function

We estimated the tip work function, $\Phi_t$, for the tip used to acquire the $dI/dV(\Delta z)$ spectra in Fig. 4 of the main text, by conducting $I_t(\Delta z)$ spectroscopy on the MOF at a pore region of the hBN/Cu(111) moiré pattern. Assuming a square tunnelling barrier, the tunnelling current $I_t$ as a function of changing tip-sample distance, $\Delta z$, is given by:

$$I_t(\Delta z) = I_0 \exp\left(-\kappa \Delta z\right), \tag{4}$$

where $I_0$ represents the current at an initial tip-sample distance $z_0$, and

$$\kappa = 2\sqrt{\frac{2m_e \Phi_{av}}{\hbar^2}}, \tag{5}$$

where $m_e$ is the electron mass, $\hbar$ the reduced Planck's constant and $\Phi_{av}$ represents the average tunneling barrier given by:

$$\Phi_{av} = \frac{\Phi_t + \Phi_s - eV_b}{2}, \tag{6}$$

where $\Phi_t$ and $\Phi_s$ are the tip and sample work functions, respectively, $V_b$ is the applied bias voltage, and $e$ the charge of an electron.

The moiré pattern of hBN/Cu(111) is known to have a maximum work function modulation of ~0.3 eV, with a work function of ~4.10 eV at the wire region and ~3.80 eV at the pore region centre [16–18]. This work function modulation depends on the specific moiré periodicity $\lambda$, which for Figs. 3 and 4 of the main text is $\lambda \approx 12.5$ nm, resulting in a work function modulation of ~0.2 eV [3]. Based on this, we estimate that, for the specific domain of interest, the MOF/hBN/Cu(111) system has an average work function of 3.95 eV (a value also consistent with DFT findings [6]), with $\Phi_s = 3.85$ eV at the pore region centre and $\Phi_s = 4.05$ eV at the wire region. Using this value for the pore region centre, we extracted $\kappa$ from the slope of $\ln[I_t(\Delta z)/I_0] = -\kappa \Delta z$ in Supplementary Fig. 25, which allowed us to estimate a tip work function of $\Phi_t = 4.84$ eV.

Alternatively, from the fits in Fig. 4e of the main text, we extracted a difference in work function between tip and sample at the wire region of 0.36 eV. Taking $\Phi_s = 4.05$ eV at the wire region, this implies a value of $\Phi_t = 4.41$ eV. The two estimates of $\Phi_t$ differ slightly, and we emphasise that they should be treated as approximate values. Importantly, however, both imply that $\Phi_t > \Phi_s$ at both the pore and wire regions of the moiré pattern, consistent with the $\Delta z$-dependence of the observed tip-induced gating in Fig. 4 of the main text and Supplementary Fig. 27.

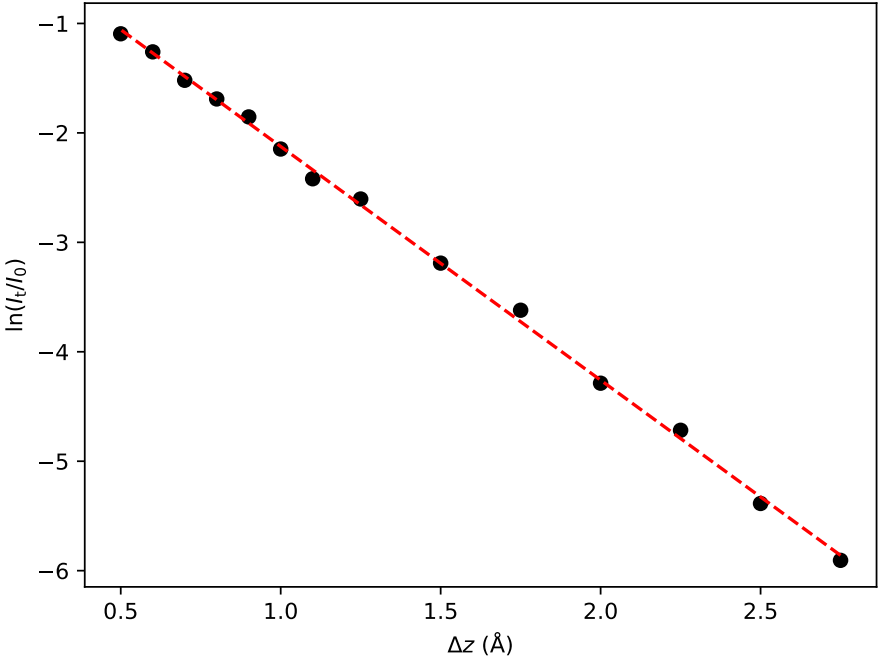

**Supplementary Fig. 25**: $\ln(I_{\mathrm{t}}/I_0)$ as a function of tip-sample distance change $\Delta z$, performed at a DCA anthracene end site of DCA$_3$Cu$_2$ MOF on hBN/Cu(111) at a bias of $V_{\mathrm{b}} = 10$ mV (pore region; hBN/Cu(111) moiré pattern periodicity $\lambda = 12.5$ nm; see red marker in Supplementary Fig. 27c for location). Dashed red line: fit using Supplementary Eq. (4), allowing for the extraction of $\kappa$. STM setpoint: $V_{\mathrm{b}} = 10$ mV, $I_0 = 10$ pA, determining initial tip-sample distance $\Delta z = 0$ Å.

# Supplementary Note 19 DBTJ model: energy diagrams

In Fig. 4 of the main text and Supplementary Fig. 26, we show schematics and energy level diagrams of the double-barrier tunnel junction (DBTJ), consisting of tip, vacuum tunnelling barrier between tip and MOF, MOF, hBN tunnelling barrier between MOF and Cu(111) substrate, and Cu(111) substrate, at different tip-sample distances $\Delta z + z_0$ and bias voltages $V_b$. With $\Phi_t > \Phi_s$ (see above), the effective electron filling of the kagome bands of the MOF decreases as $\Delta z$ decreases. This process sequentially drives the Mott insulator to metal to trivial insulator transition. The MOF can become charged due to resonant electron transfer when MOF electronic states align with the Cu(111) substrate Fermi level.

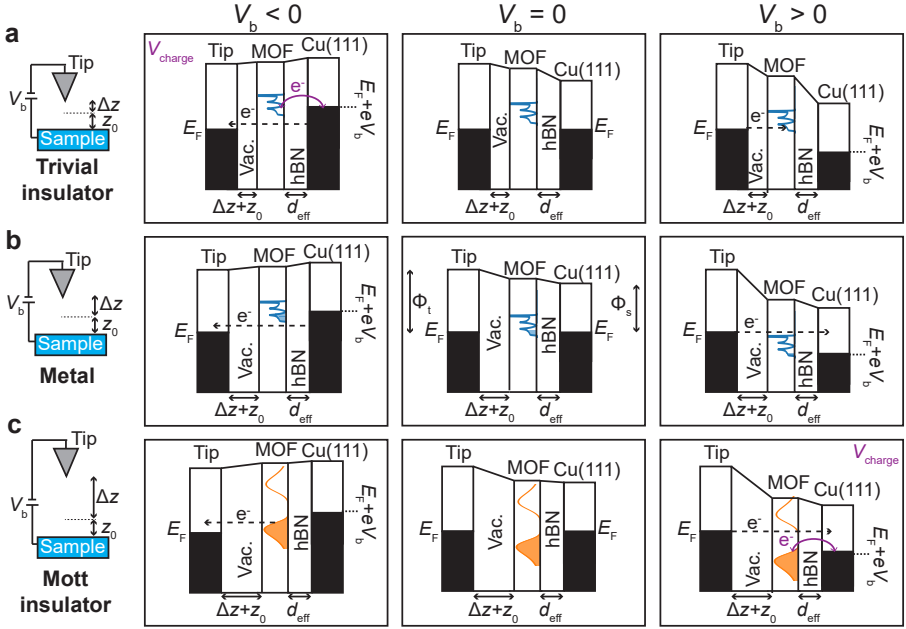

**Supplementary Fig. 26**: **Double-barrier tunnelling junction (DBTJ) model of tip-induced gating of MOF at wire region of hBN/Cu(111) moiré pattern. a-c,** Schematics and energy diagrams of tip/vacuum/-MOF/hBN/Cu(111) DBTJ, for small, intermediate and large tip-sample distances $\Delta z + z_0$. When $V_b$ is applied to the Cu(111) substrate (with tip grounded), a voltage drop occurs at the MOF location, resulting in MOF energy level shifts. These energy shifts can lead to MOF charging and to effective emptying/filling of MOF electronic states, allowing for gating of MOF transitions, from (correlated) Mott insulator to metal to trivial insulator. Some of these panels are shown in Fig. 4d of the main text. These schematics are qualitative representations of the proposed mechanism of tip-induced gating.

Here, we provide more detailed information on this tip-induced transition from Mott insulator to metal to trivial insulator, and on the charging of MOF electronic states.

The $dI/dV$ for the MOF at the hBN/Cu(111) moiré wire region in Fig. 4d of the main text show an electronic gap, with a clear peak (sharper than the near-Fermi band features in Fig. 3b, with a maximum indicated by purple circles) at positive bias voltage for large tip-sample distances, and at negative bias voltage for small tip-sample distances (see also Supplementary Fig. 28). These spectra also show a subtler band edge (indicated by the red squares in main text Fig. 4d, similar to the near-Fermi band features in Fig. 3b) at a bias voltage of sign opposite to that of the sharp peak – at negative bias voltage for large tip-sample distances and at positive bias voltage for small

tip-sample distances. For intermediate tip-sample distances, these spectra are gapless, with a significantly larger Fermi-level $dI/dV$ signal (see main text Fig. 4f). From these observations, we infer that the MOF is an insulator (trivial or Mott) for large and small tip-sample distances ($dI/dV$ spectra with a gap at the Fermi level), and a metal for intermediate tip-sample distances ($dI/dV$ spectra with no gap at the Fermi level). The bias voltage position of the sharp peak (purple circles in Fig. 4d) increases linearly with respect to tip-sample distance, whereas the bias voltage position of the subtler band edge decreases non-linearly with tip-sample distance. Notably, DBTJ model Eqs. (1) and (2) in the main text provide very good fits for the bias voltage positions of both sharp peak and subtle band edge as a function of tip-sample distance (black curves in main text Fig. 4e). This provides compelling evidence that the sharp peaks (purple circles) are associated with charging of an intrinsic MOF electronic state lying at the edge (red squares) of a populated (large tip-sample distances) or empty (small tip-sample distances) band (see main text Fig. 4a-c and Supplementary Fig. 26).

In Fig. 2 of the main text, we show that the $dI/dV$ spectra for the MOF at a pore region of the hBN/Cu(111) moiré pattern show a $\sim$200 meV electronic energy gap at the Fermi level. These spectra resemble the spectral function of the 2D kagome MOF in the Mott insulating phase (main text Fig. 1e), calculated via DMFT with a chemical potential that is consistent with the DFT-predicted occupation of the near-Fermi kagome bands for the MOF on hBN/Cu(111) (main text Fig. 1d, Supplementary Fig. 2). At an adjacent moiré wire region, the local work function increases by $\sim$0.2 eV for the specific periodicity of the MOF/hBN/Cu(111) domain considered in main text Fig. 3. Accordingly, the near-Fermi electronic states of the MOF at this moiré wire region are shifted upwards in energy in comparison to the near-Fermi electronic states of the MOF at the moiré pore region. The $dI/dV$ spectra in Fig. 3b for this moiré wire region (for the specific tunnelling parameters used) show no gap at the Fermi level, with a significant non-zero Fermi-level $dI/dV$ signal (main text Fig. 3e), indicative of a metallic phase. These experimental $dI/dV$ spectra are consistent with the spectral function $A(E)$ of the MOF calculated via DMFT for a chemical potential that is reduced (in comparison with the DMFT calculations for the moiré pore region). This DMFT spectral function exhibits a significant magnitude and no gap at the Fermi level, indicating a metallic phase, resulting from the depopulation of the MOF bands (main text Fig. 3d).

From these observations we infer that: (i) the MOF at the moiré pore region is in a Mott insulating phase, and (ii) the MOF at the adjacent moiré wire region, for the specific tunnelling parameters used (bias voltage, tip-sample distance) in main text Fig. 3b, is in a metallic phase (as the result of the depopulation of the MOF near-Fermi electronic states due to the increase in local work function).

The work function of the STM tip is larger than that of the sample at the moiré wire region (Supplementary Note 18). When the tip-sample distance is reduced, the DBTJ leads to an upward energy shift of the MOF electronic states (with respect to the Cu(111) Fermi level): the MOF electronic states become further depopulated (Supplementary Fig. 26). That is, when the $dI/dV$ spectra in main text Fig. 4d transition from gapless (metallic) to gapped (insulator) as the tip-sample distance decreases, we infer that the MOF electronic states associated with the MOF kagome bands become empty, with the Fermi level lying below the bottom of the three near-Fermi kagome bands (Supplementary Fig. 1). From this, we associate the gapped $dI/d$ spectra at the top of main text Fig. 4d (for small tip-sample distances) to a trivial insulating phase of the MOF.

By contrast, when the tip-sample distance is increased (large $\Delta z$ regime in Fig. 4 of main text), the DBTJ leads to a downward energy shift of MOF electronic states [with respect to the Cu(111) Fermi level], which become more populated (Supplementary Fig. 26). When the $dI/dV$ spectra in main text Fig. 4d transition from gapless (metallic) to gapped (insulator) as the tip-sample distance increases, we infer that the population of the MOF electronic states associated with the MOF kagome bands also increases in turn, reaching a threshold (half-filling of the bands) that opens the Mott gap, with the Fermi lying in such Mott gap. This population of MOF states (here driven by the DBTJ effect and the increase in tip-sample distance) is analogous to the population of MOF states at the adjacent moiré pore region (due to the smaller local work function at such a moiré pore region; main text Fig. 3c). From this, we associate the gapped $dI/dV$ spectra at the bottom of main text Fig. 4d (for large tip-sample distances) to the Mott insulating phase of the MOF.

This inference of a Mott gap at the moiré wire region for large tip-sample distances is supported by DMFT. Indeed, when the chemical potentials used in the DMFT calculations of main text Fig. 3d are all offset upwards by 45 meV (leading to further population of the MOF electronic states, mimicking the effect of a tip-sample distance increase), all spectra lie in the Mott insulating regime with no evidence of metallic character (Supplementary Note 4).

# Supplementary Note 20  Tip-induced gating at moiré pore region

In addition to the tip-sample distance dependent $dI/dV$ spectra shown in Fig. 4 of the main text for the wire region of the hBN/Cu(111) moiré pattern, we performed equivalent measurements at the pore region. These are shown in Supplementary Fig. 27. All these spectra exhibit an energy gap for all values of tip-sample distance $\Delta z + z_0$ considered, with no indication of tip-induced MOF charging or of Mott metal-insulator transition. We infer that, at this pore site, the MOF remains robustly within the Mott insulating phase for all tip-sample distances explored.

These $dI/dV$ spectra do show however small variations of the lower Hubbard band maximum (LHBM), $V_{LHB}$, and upper Hubbard band minimum (UHBM), $V_{UHB}$, as a function of $\Delta z$ (Supplementary Fig. 27b). According to the DBTJ model [see Eq. (1) in the main text], this variation should follow:

$$V_{LHB}(\Delta z) = \frac{d_{eff}\left(V_{\infty}^{(LHB)} + \Delta\Phi_{ts}\right)}{(\Delta z + z_0)} + V_{\infty}^{(LHB)}, \qquad (7)$$

$$V_{UHB}(\Delta z) = \frac{d_{eff}\left(V_{\infty}^{(UHB)} + \Delta\Phi_{ts}\right)}{(\Delta z + z_0)} + V_{\infty}^{(UHB)}, \qquad (8)$$

where $d_{eff}$ is the effective width of the hBN tunnel barrier, $\Delta z + z_0$ is the width of the vacuum tunnel barrier (tip-MOF distance, with $z_0$ representing an initial tip-sample distance with respect to an STM setpoint), $V_{\infty}^{(LHB)}$ ($V_{\infty}^{(UHB)}$) is the bias voltage corresponding to the LHBM (UHBM, respectively) as $\Delta z \to \infty$, and $\Delta\Phi_{ts}$ is the difference between tip and sample work functions [15].

We fit $V_{LHB}(\Delta z)$ and $V_{UHB}(\Delta z)$ according to Supplementary Eqs. (7), (8), by using a global fitting approach which maintained the same values of $d_{eff}$, $z_0$ and $\Delta\Phi_{ts}$ for both fits (dashed black lines in Supplementary Fig. 27b). These fits show good agreement between the experimental data and the DBTJ model.

The change in energy of the LHBM and UHBM with $\Delta z$ is more gradual at the pore site than the LHBM at the wire site (see Fig. 4 in the main text). While the difference between tip and sample work functions is expected to be larger at the pore site than at the wire (see Supplementary Note 18), this may be explained by a smaller $d_{eff}$ at the pore site than at the wire site. For our system, $d_{eff} = d_{hBN}/\epsilon$, where $d_{hBN}$ is the MOF-Cu(111) distance ($\sim$hBN thickness) and $\epsilon$ is the out-of-plane relative dielectric constant associated with the hBN tunnel barrier [15]. From the fits in Supplementary Fig. 27c and Fig. 4e of the main text, we find $d_{eff}^{(pore)} = 0.06 \pm 0.05$ Å and $d_{eff}^{(wire)} = 0.2 \pm 0.1$ Å. While the hBN/Cu(111) moiré pattern is predominantly an electronic effect, some estimates of the structural corrugation between pore and wire sites are as large as 0.7 Å [19] – consistent with our findings for $d_{eff}$. This suggests that within the DBTJ model, the MOF at the wire regions of the hBN/Cu(111) moiré pattern may be more decoupled from the underlying Cu(111) than the MOF at the pore regions.

Additionally, no evidence of charging features were observed within the measurements acquired for the MOF at the pore sites of the hBN/Cu(111) moiré pattern. This can be explained by a larger energy difference between LHBM and $E_F$ at the pore regions compared to the wire regions, meaning that greater values of $V_b$ are required to induce charging.

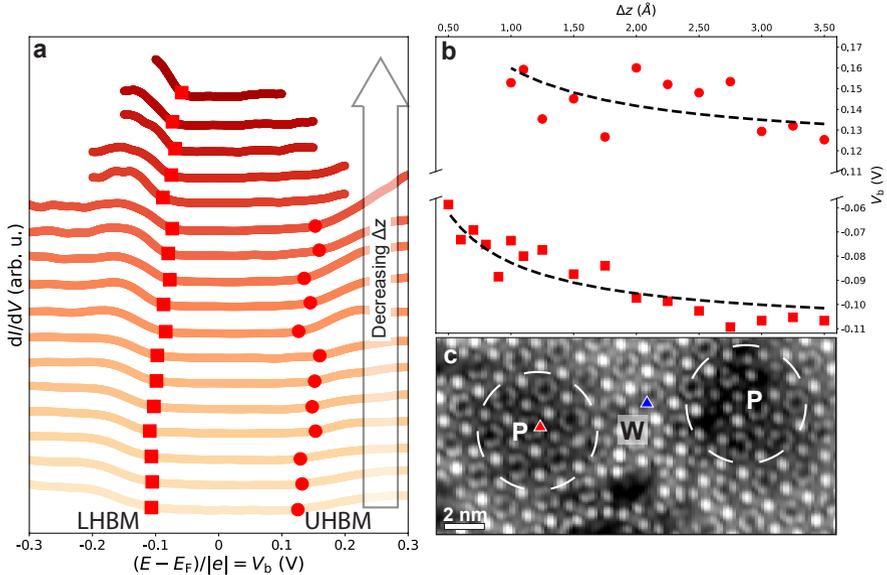

**Supplementary Fig. 27**: d$I$/d$V$ spectroscopy for **DCA$_3$Cu$_2$ MOF** at the centre of a pore region of hBN/Cu(111) moiré pattern. **a,** d$I$/d$V$ spectra for different tip-sample distances $\Delta z + z_0$ ($z_0$ represents tip-sample distance for STM setpoint of $V_b = 10$ mV, $I_t = 10$ pA) at a DCA lobe site (location indicated by red triangle in c). Red squares (circles) indicate LHBM (UHBM). **b,** LHBM and UHBM as a function of $\Delta z$, from (a). Black dashed curves: fits based on Supplementary Eqs. (7) and (8). **c,** STM image showing DCA$_3$Cu$_2$ MOF at pore and wire regions of hBN/Cu(111) moiré pattern. Red triangle indicates position where spectra shown in (a) were acquired; blue triangle indicates position where spectra shown in Fig. 4 of main text were acquired ($V_b = -1$ V, $I_t = 10$ pA).

# Supplementary Note 21 Tip-induced gating at Cu sites; DBTJ model fit parameters

In Fig. 4 and Supplementary Fig. 27, we showed tip-induced gating of the MOF at a DCA lobe site located at a wire and at a pore region of the hBN/Cu(111) moiré pattern (with moiré period $\lambda \approx 12.5$ nm), respectively. Here, we demonstrate that measurements performed at MOF Cu sites are similar and consistent (Supplementary Figs. 28, 29).

Supplementary Fig. 28 shows d$I$/d$V$ spectra for different tip-sample distances $z_0 + \Delta z$, performed on a Cu site of the DCA$_3$Cu$_2$ MOF at a wire region of the hBN/Cu(111) moiré pattern (with moiré periodicity $\lambda \approx 12.5$ nm, as in the main text). As in Fig. 4, the spectra acquired at this site feature, for large $\Delta z$, at negative $V_b$, a LHB whose energy maximum (LBHM)

increases (approaches the Fermi level) with decreasing $\Delta z$. These spectra also exhibit, at positive $V_{\mathrm{b}}$, a peak whose energy position decreases linearly (and more abruptly than the LHBM) with decreasing $\Delta z$. We tracked the LHBM (assigned using the method outlined in Supplementary Note 13) and the positive bias peak (assigned as a local maximum in $\mathrm{d}I/\mathrm{d}V$) as a function of $\Delta z$, and then fit both features using Eqs. (1) and (2) of the main text, respectively, via a global fitting approach (see also main text Methods section 'DBTJ model'). The strong agreement between fits and data in Supplementary Fig. 28b shows that the DBTJ model captures the experimental behaviour well, with the positive bias peak corresponding to the charging of electronic states at the LHBM, as for Fig. 4. Similar to Fig. 4, as $\Delta z$ is further decreased (for $\Delta z < 1.2$ Å), we observe a significant increase in Fermi-level ($V_{\mathrm{b}} = 0$) $\mathrm{d}I/\mathrm{d}V$ signal. As for the DCA lobe site in Fig. 4 of the main text, we interpret this as the transition from Mott insulator to metal via tip-induced gating (combined with the influence of the local work function at this moiré wire region). Within the range of experimental variables explored, the transition to a trivial insulator (see top curves in Fig. 4d of main paper) was not reached for this particular Cu site dataset.

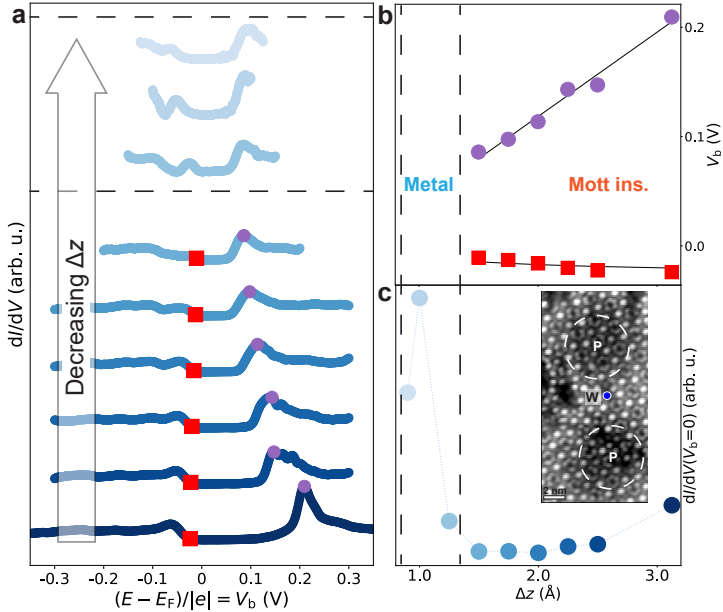

**Supplementary Fig. 28**: **Tip-induced gating at a Cu site of the MOF within a wire region of the hBN/Cu(111) moiré pattern. a,** $dI/dV$ spectra at MOF Cu site, for different $\Delta z + z_0$ ($z_0$ given by STM setpoint $V_b = 10$ mV, $I_t = 10$ pA). Purple circles (red squares): MOF charging peak (intrinsic electronic state at MOF band edge, respectively). Spectra normalised and offset for clarity. **b,** $V_{charge}$ [purple circles in (a)] and $V_{state}$ [red squares in (a)] as a function of $\Delta z$. Black solid lines: global fits to Eqs. (1) and (2) in main text. **c,** $dI/dV$ signal at Fermi level ($V_b = 0$) as a function of $\Delta z$, from (a). Increased $dI/dV(V_b = 0)$ indicates metallic phase. Inset: STM image with position where $dI/dV(\Delta z)$ were performed indicated by blue circle ($V_b = -1$ V, $I_t = 10$ pA).

Similarly, Supplementary Fig. 29a shows $dI/dV$ spectra for different tip-sample distances $z_0 + \Delta z$, performed on a Cu site of the $DCA_3Cu_2$ MOF at a pore region of the hBN/Cu(111) moiré pattern (with moiré periodicity $\lambda \approx 12.5$ nm, as in the main text). Consistent with Supplementary Fig. 27 for measurements at a DCA lobe site, Supplementary Fig. 29b shows the evolution of LHBM and UHBM as a function of $\Delta z$, well described by the DBTJ model Supplementary Eqs. (7) and (8) (dashed curves), respectively.

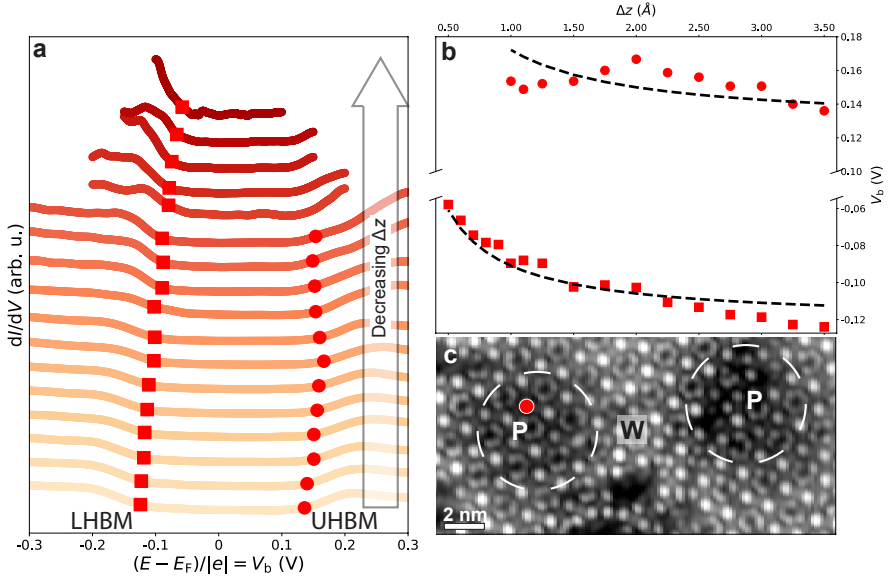

**Supplementary Fig. 29**: **d$I$/d$V$ spectroscopy for Cu site of MOF within a pore region of hBN/Cu(111) moiré pattern. a,** d$I$/d$V$ spectra for different tip-sample distances $\Delta z + z_0$ ($z_0$ represents tip-sample distance for STM setpoint of $V_b = 10$ mV, $I_t = 10$ pA) at a Cu site [location indicated by red circle in (c)]. Red squares (circles) indicate LHBM (UHBM, respectively). **b,** LHBM and UHBM as a function of $\Delta z$, from (a). Black dashed curves: fits based on Supplementary Eqs. (7) and (8). **c,** STM image of DCA$_3$Cu$_2$ MOF on hBN/Cu(111). Red circle indicates position where spectra shown in (a) were acquired ($V_b = -1$ V, $I_t = 10$ pA).

The fitting parameters corresponding to the fits in Fig. 4e (main text), and Supplementary Figs. 27b, 28b, and 29b, are summarised in Supplementary Table 1. Equations (1) and (2) of the main text were used to fit the spectra acquired at wire regions of the hBN/Cu(111) moiré pattern in Fig. 4e of the main text (bottom curves) and Supplementary Fig. 28b (and exhibiting a gap at the Fermi level), with fitting parameters $V_\infty^{(\text{Mott})}$, $d_{\text{eff}}$, $\Delta\Phi_{\text{ts}}$, and $z_0$. For the top spectra in Fig. 4e of the main text (that we attribute to the trivial insulator phase), we used the same two equations with fitting parameter $V_\infty^{(\text{kagome})}$ corresponding to the energy level at the bottom of the near-Fermi kagome bands. Supplementary Eqs. (7) and (8) were used to fit spectra acquired at the pore regions of the hBN/Cu(111) moiré pattern in Supplementary Figs. 27 and 29 with fitting parameters $V_\infty^{(\text{LHB})}$, $V_\infty^{(\text{UHB})}$, $d_{\text{eff}}$, $\Delta\Phi_{\text{ts}}$, and $z_0$.

The fitting parameters $V_\infty^{(\text{Mott/LHB})}$, $V_\infty^{(\text{UHB})}$ and $d_{\text{eff}}$ – characteristic of the MOF system on hBN/Cu(111) – obtained for the fit curves in Fig. 4e of the main text, and Supplementary Figs. 27b, 28b for DCA lobe and Cu

sites are highly consistent (within uncertainty), for both pore and wire regions of the hBN/Cu(111) moiré pattern. The global fitting approach used does lead to large uncertainties (determined from fitting procedure as one standard deviation of fit parameters) in some cases, however, notably for values of $z_0$. As discussed in the previous Supplementary Note 20, the difference in $d_{eff}$ between pore and wire regions is highly consistent with hBN/Cu(111) literature and supports our claim that the MOF is slightly more decoupled at the wire regions than at the pore regions [19]. Importantly, for data acquired with the same tip (DCA lobe measurements at both pore and wire regions, and Cu measurements at the pore region), $\Delta\Phi_{ts}^{(pore)} - \Delta\Phi_{ts}^{(wire)} = (\Phi_t - \Phi_{pore}) - (\Phi_t - \Phi_{wire}) = \Phi_{wire} - \Phi_{pore} \approx 0.2$ eV, consistent with the expected local work function corrugation of the hBN/Cu(111) moiré pattern of interest with $\lambda \approx 12.5$ nm. Similarly, the difference in $V_\infty^{(Mott/LHB)}$ at pore and wire regions – with $V_\infty^{(Mott/LHB)}$ significantly closer to the Fermi level at the wire region than at the pore region – is consistent with the findings shown in Fig. 3 of the main text and Supplementary Figs. 19-22. Finally, it is important to note that the condition of the STM tip was different for measurements at the Cu site in a moiré wire region to the other three datasets, providing a potential explanation for the different values of $\Delta\Phi_{ts}$ and $z_0$. Nonetheless, as for the other datasets, a $\Delta\Phi_{ts} > 0$ was obtained from the fit at such wire region Cu site, indicating a tip-induced gating behaviour consistent with other measurements. This provides further evidence that the DBTJ model quantitatively captures the observed changes in $dI/dV$ measurements with changing tip-sample distance.

| Quantity | DCA lobe sites | | Cu sites | |
|---|---|---|---|---|
| | Pore | Wire | Pore | Wire* |
| $V_\infty^{(Mott/LHB)}$ (V) | -0.11 ± 0.01 | -0.03 ± 0.02 | -0.12 ± 0.02 | -0.03 ± 0.02 |
| $V_\infty^{(kagome)}$ (V) | | -0.02 ± 0.01 | | |
| $V_\infty^{(UHB)}$ (V) | 0.12 ± 0.01 | | 0.13 ± 0.03 | |
| $d_{eff}$ (Å) | 0.06 ± 0.05 | 0.2 ± 0.1 | 0.06 ± 0.08 | 0.4 ± 0.3 |
| $\Delta\Phi_{ts}$ (eV) | 0.6 ± 0.4 | 0.36 ± 0.09 | 0.6 ± 0.5 | 0.1 ± 0.2 |
| $z_0$ (Å) | 0.2 ± 0.6 | 0.007 ± 0.5 | 0.03 ± 1 | 0.4 ± 2 |

**Supplementary Table 1: DBTJ model fitting parameters.** Eqs. (1) and (2) in main text were used to fit the $dI/dV$ spectra acquired at wire regions of the hBN/Cu(111) moiré pattern in Fig. 4e (main text) and Supplementary Fig. 28b. Supplementary Eqs. (7) and (8) were used to fit spectra acquired at the pore regions of the hBN/Cu(111) moiré pattern in Supplementary Figs. 27 and 29. *The Cu site dataset at the moiré wire region was acquired with a different STM tip to the other three datasets.

# Supplementary Note 22 Tip-induced gating: evidence of metallic character of MOF at wire region

In Fig. 4 of the main text, we show that the MOF undergoes a transition from a Mott insulating phase to a metal phase to a trivial insulator phase via tip-induced gating at the wire region of the hBN/Cu(111) moiré pattern. The metallic phase is evidenced by the absence of an energy gap at the Fermi level, for a small window of intermediate tip-sample distances $\Delta z + z_0$. Notably, the $dI/dV$ spectra acquired at these distances have increased $dI/dV(V_b = 0)$ signal; see Supplementary Fig. 30b-c. This is indicative of an increased spectral function at the Fermi level, and hence of metallic character.

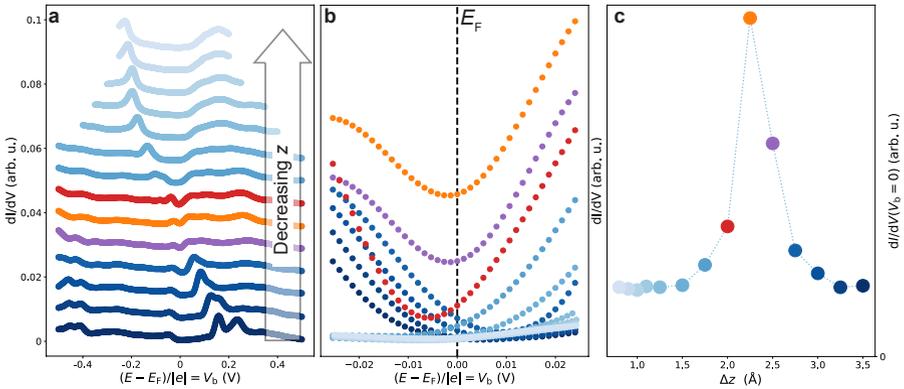

**Supplementary Fig. 30**: **Evidence of tip-induced metallic character of DCA$_3$Cu$_2$ MOF at wire region of hBN/Cu(111) moiré pattern. a,** Normalised $dI/dV(\Delta z)$ spectra (also shown in Fig. 4 of main text) acquired at DCA anthracene extremity site at wire region, for different tip-sample distances $\Delta z$ ($\Delta z = 0$ corresponds to STM setpoint of $V_b = 10$ mV, $I_t = 10$ pA). Curves offset for clarity. **b,** Same $dI/dV(\Delta z)$ spectra as in (a), for near-Fermi energy range (no offset). Purple, orange and red spectra for intermediate $\Delta z + z_0$ show increased $dI/dV$ signal at Fermi level ($V_b = 0$), suggesting metallic character with no gap at the Fermi level. Vertical black dashed line: Fermi level. **c,** Value of $dI/dV$ at $E_F$ ($V_b = 0$) as a function of $\Delta z$, illustrating metallic character for the purple, orange and red points.

# Supplementary Note 23 Temperature-dependent measurements and calculations

To assess the influence of temperature on the electronic properties of the $DCA_3Cu_2$ MOF, we conducted further STS measurements at both pore and wire sites of the hBN/Cu(111) moiré pattern, at a temperature of 77 K (Supplementary Fig. 31a). Supplementary Fig. 31b shows measurements at 4 K, for comparison. While the 77 K spectra appear broader than the 4 K spectra (due to thermal broadening), they qualitatively show the same electronic features: metallic behaviour at the wire region (blue, with no gap at the Fermi level and a larger Fermi-level $dI/dV$ signal), and insulating behaviour at the pore region (orange, with an energy gap at the Fermi level).

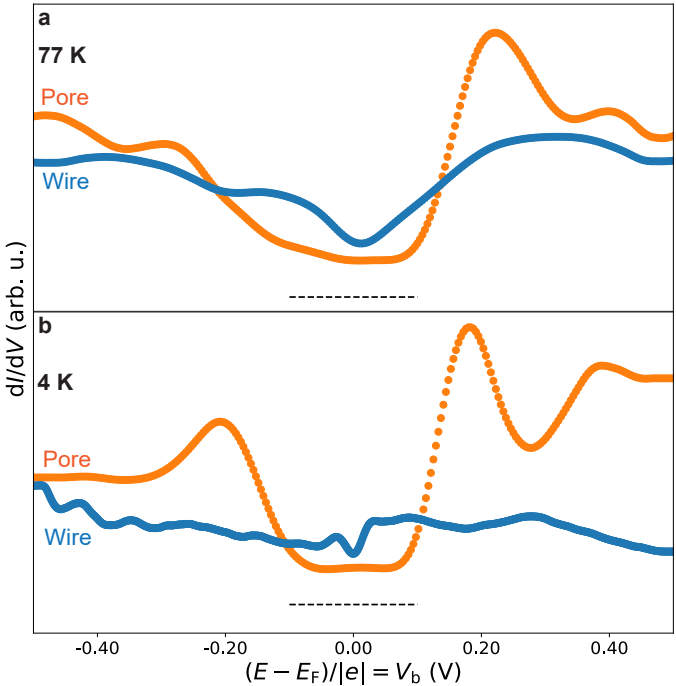

**Supplementary Fig. 31**: **Temperature-dependent STS measurements.** $dI/dV$ measurements performed at DCA lobe sites of the $DCA_3Cu_2$ MOF at hBN/Cu(111) moiré pore (orange) and wire (blue) regions, at temperatures of **(a)** 77 K and **(b)** 4 K. Setpoints: (a) $V_b = -500$ mV, $I_t = 100$ pA; (b) $V_b = -500$ mV, $I_t = 500$ pA (orange), and $V_b = -500$ mV, $I_t = 160$ pA (blue). Black dashed horizontal lines indicate position of $dI/dV = 0$.

We also explored temperature-dependence using DMFT. Supplementary Fig. 32 shows the evolution of DMFT spectral functions, $A(E)$, for different temperatures, with $U = 0.65$ eV and a chemical potential $E_F$ corresponding to the pore (Supplementary Fig. 32a) or wire regions (Supplementary Fig. 32b) of the hBN/Cu(111) moiré pattern. DMFT predicts no qualitative change in the MOF's electronic properties as the temperature evolves from 29 K to beyond room temperature, consistent with the experimental observations in Supplementary Fig. 31.

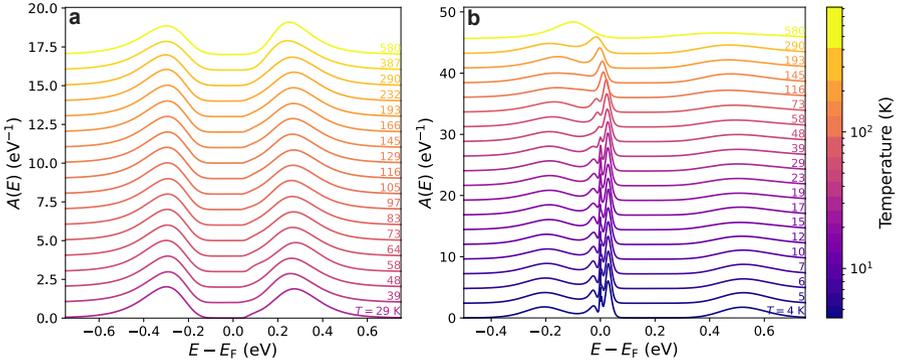

**Supplementary Fig. 32**: **Temperature-dependent spectral functions calculated by DMFT. a,** $E_F = 0.4$ eV (∼half-filling, corresponding to moiré pore regions). **b,** $E_F = 0.2$ eV (filling corresponding to metal-like phase observed experimentally at wire regions). $U = 0.65$ eV.

# Supplementary Note 24 STS measurements at different moiré pore regions

To examine how consistent the electronic properties of the DCA$_3$Cu$_2$ MOF are across the surface, we acquired d$I$/d$V$ spectra at Cu and DCA lobe MOF sites, at different moiré pore regions within a hBN/Cu(111) domain with moiré period $\lambda \approx 12.5$ nm (Supplementary Fig. 33). These spectra are very similar to each other, showing LHB and UHB signatures, and a gap $E_g \approx 0.2$ eV at the Fermi level. Some subtle differences (e.g., shifting in energy of Hubbard bands) may be attributed to spectra acquired at slightly different distances from the moiré pore centre (hence with a slightly different hBN/Cu(111) local work function), or to different setpoints (see Supplementary Note 20). This shows the consistency of the MOF electronic properties at equivalent sites within a hBN/Cu(111) domain.

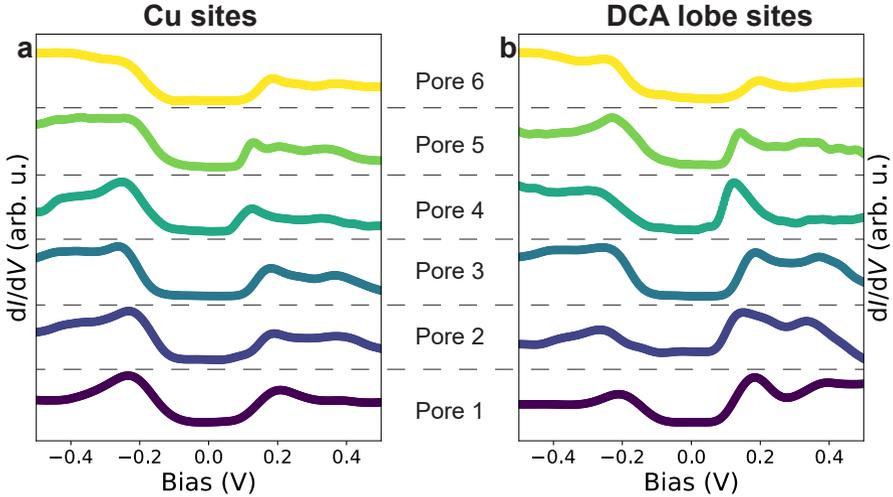

**Supplementary Fig. 33**: d$I$/d$V$ **spectra at different moiré pore regions within hBN/Cu(111) domain with moiré period $\lambda \approx 12.5$ nm. a,** Cu MOF sites. **b,** DCA lobe sites. Acquisition sites close to centre of moiré pore regions. Spectra normalised and offset for clarity. Pore 1 setpoint: $V_b = -500$ mV, $I_t = 500$ pA. Pores 2,3 setpoints: 190 pm further from surface with respect to setpoint of $V_b = 10$ mV, $I_t = 10$ pA. Pores 4,5 setpoint: 225 pm further from surface with respect to setpoint of $V_b = 10$ mV, $I_t = 10$ pA. Pore 6 setpoint: 150 pm further from surface with respect to setpoint of $V_b = 10$ mV, $I_t = 10$ pA.



\end{document}